\documentclass[superscriptaddress,twocolumn,showpacs,preprintnumbers,amsmath,amssymb,nofootinbib,eqsecnum,longbibliography]{revtex4-1}
\usepackage{graphicx}

\begin{document}

\title{Deformed and twisted black holes with NUTs}

\author{Pavel Krtou\v{s}}

\email{Pavel.Krtous@utf.mff.cuni.cz}

\affiliation{Institute of Theoretical Physics,
Faculty of Mathematics and Physics, Charles University,
V~Hole\v{s}ovi\v{c}k\'ach~2, Prague, 18000, Czech Republic}

\author{David Kubiz\v n\'ak}

\email{dkubiznak@perimeterinstitute.ca}

\affiliation{Perimeter Institute, 31 Caroline Street North, Waterloo, ON, N2L 2Y5, Canada}


\author{Valeri P. Frolov}

\email{frolov@phys.ualberta.ca}

\affiliation{Theoretical Physics Institute, University of Alberta, Edmonton, Alberta, Canada T6G 2G7}

\author{Ivan Kol\'{a}\v{r}}

\email{Ivan.Kolar@utf.mff.cuni.cz}

\affiliation{Institute of Theoretical Physics,
Faculty of Mathematics and Physics, Charles University,
V~Hole\v{s}ovi\v{c}k\'ach~2, Prague, 18000, Czech Republic}

\affiliation{Theoretical Physics Institute, University of Alberta, Edmonton, Alberta, Canada T6G 2G7}

\date{November 8, 2015}   

\begin{abstract}
We construct a new class of vacuum black hole solutions whose geometry is deformed and twisted by the presence of NUT charges. The solutions are obtained by `unspinning' the general Kerr-NUT-(A)dS spacetimes, effectively switching off some of their rotation parameters. The resulting geometry has a structure of warped space with the Kerr-like Lorentzian part warped to a Euclidean metric of deformed and/or twisted sphere, with the deformation and twist characterized by the `Euclidean NUT' parameters. In the absence of NUTs, the solution reduces to a well known Kerr-(A)dS black hole with several rotations switched off. New geometries inherit the original symmetry of the Kerr-NUT-(A)dS family, namely, they possess the full Killing tower of hidden and explicit symmetries. As expected, for vanishing NUT, twist, and deformation parameters, the symmetry is further enlarged.
\end{abstract}


\maketitle

\section{Introduction}

In four dimensions, a large family of solutions of vacuum Einstein equations with the cosmological constant of type D have been constructed by Pleba\'{n}ski and Demia\'{n}ski \cite{plebanski1976rotating}, generalizing the Carter--Pleba\'{n}ski form \cite{Carter:1968b, plebanski1975class} of non-accelerating solutions. Various known metrics, for example the Kerr metric \cite{Kerr:1963}, the Taub-NUT solution \cite{taub1951empty, newman1963empty, misner1963flatter, Clement:2015cxa}, or the (accelerating) C-metric e.g. \cite{stephani2003exact} are contained in this family as `special cases'. However, to obtain such special cases explicitly is not an easy task, as one has to perform certain (often singular) limits, scaling both the coordinates and the metric parameters. It is also for this reason the physical meaning of parameters of the Pleba\'{n}ski--Demia\'{n}ski solution, describing for example the rotation or a NUT charge, is often obscure and difficult to interpret, see e.g. \cite{GriffithsPodolsky:2006b, GriffithsPodolsky:book}. On the other hand, the Carter--Pleba\'{n}ski frame is directly linked to the underlying explicit and hidden symmetries of the geometry and hence is invaluable for the study of mathematical properties of the spacetime. For example, it is in this frame the test scalar field equations do separate \cite{Carter:1968b}. Moreover, being able to describe a large family of spacetimes, the Carter--Pleba\'{n}ski (or more generally Pleba\'{n}ski--Demia\'{n}ski) metric allows one to generate (by taking special limits) new types of solutions, see, e.g., \cite{Hubeny:2009kz, Gnecchi:2013mja, Nozawa:2015qea} for recent examples.

Although a generalization of the Pleba\'{n}ski--Demia\'{n}ski family to higher dimensions remains elusive (see however \cite{Lu:2008js}), a higher-dimensional generalization of the Carter--Pleba\'{n}ski class, the so-called \emph{Kerr-NUT-(A)dS spacetime}, has been found by Chen, L\"{u}, and Pope \cite{ChenLuPope:2006}. The geometry describes {\em a large family  of metrics} of various signatures that solve the Einstein equations with the cosmological constant and contain a set of free parameters that can be related to mass, rotations, and NUT parameters. In particular, the general rotating black holes of Myers and Perry \cite{MyersPerry:1986}, their cosmological constant generalizations due to Gibbons et al. \cite{GibbonsEtal:2004,GibbonsEtal:2005}, or the higher-dimensional Taub-NUT spaces e.g. \cite{Mann:2003zh, Mann:2005ra, Clarkson:2006zk, ChenLuPope:2007}, are expected to emerge as certain limits of Kerr-NUT-(A)dS spacetimes.

Surprisingly, apart from its `integrability properties' not much is known about the Kerr-NUT-(A)dS geometry and its special subcases. On one hand, it has been shown \cite{Houri:2007xz, KrtousFrolovKubiznak:2008} that, similar to the four-dimensional Carter--Pleba\'{n}ski case, the Kerr-NUT-(A)dS spacetime represents a unique Einstein space that admits a completely non-degenerate {\em closed conformal Killing--Yano} (CCKY) 2-form $\tens{h}$ \cite{FrolovKubiznak:2007,KubiznakFrolov:2007}. This tensor uniquely determines {\em canonical coordinates} in which the metric is written. In these coordinates the Hamilton--Jacobi, Klein--Gordon, and Dirac equations fully separate \cite{FrolovEtal:2007,SergyeyevKrtous:2008,KolarKrtous:2015b,OotaYasui:2008a,Oota:2008uj,CarigliaEtal:2011a,CarigliaEtal:2011b}---making in particular the geodesic motion completely integrable \cite{PageEtal:2007,KrtousEtal:2007a, Houri:2007uq, KrtousEtal:2007b,Krtous:2015}. Such integrability properties can be further linked to the existence of a full {\em Killing tower} of explicit and hidden symmetries that can be generated from $\tens{h}$, see e.g. \cite{FrolovKubiznak:2008,Frolov:2012ux} for a review. On the other hand, again similar to the Carter--Pleba\'{n}ski case, it is not straightforward to precisely identify the exact physical meaning of the metric parameters, nor is particularly simple to obtain special subcases contained in the Kerr-NUT-(A)dS family. For example, the non-rotating Schwarzschild--Tangherlini black hole \cite{Tangherlini:1963bw} cannot be simply recovered by merely setting some of the parameters equal to zero.

The goal of this paper is to proceed towards better understanding of the higher-dimensional Kerr-NUT-(A)dS family \cite{ChenLuPope:2006} as well as to uncover some of its  special subcases. We start by performing a basic analysis of the Kerr-NUT-(A)dS spacetime, discussing its signature, coordinate ranges, scaling properties, and number of independent metric parameters. In particular, we identify the maximally symmetric space, the Euclidean instanton, and the black hole solution. Concentrating on the black hole case, we then show how to switch off some of the rotation parameters---constructing so a new rich class of vacuum (with a possible cosmological constant) black holes that can be distinguished not only by mass and rotation parameters but also by NUT charges.

More concretely, we want to study black hole spacetimes that are generated from the Kerr-NUT-(A)dS geometry in the limit where some of the rotation parameters vanish. Unfortunately, one cannot just substitute zero values of the chosen rotational parameters directly in the Kerr-NUT-(A)dS metric, since the ranges of some coordinates would degenerate. This is related to a well known ambiguity in constructing the limiting spaces when some of the parameters limit to zero \cite{Geroch:1969}: there is always a possibility to make a coordinate transformation depending on the chosen parameters, before taking the limit. To escape the pathology and to achieve the limit of vanishing rotations, we have to properly {\em rescale both} the {\em metric parameters} and the {\em coordinates}.

As we will demonstrate, the resulting black hole spacetimes have one parameter less than their Kerr-NUT-(A)dS precursor and inherit from it nice geometrical properties. In the absence of NUTs, they reduce to the Kerr-(A)dS black holes \cite{GibbonsEtal:2004,GibbonsEtal:2005} with several rotations switched off, in particular we obtain the Schwarzschild--Tangherlini-(A)dS solution \cite{Tangherlini:1963bw}. However, when the NUT parameters are maintained, the black hole spacetimes are more interesting and, we believe, new. The metric acquires a warp product structure, where both components of the product have the form similar to the Kerr-NUT-(A)dS metric. Namely, it takes the following form:
\be
\mtrc = \mtrt+\warp^2 \mtrb\,,
\ee
where the Euclidean {\em seed metric} of the product, $\mtrb$, is exactly the lower dimensional Kerr-NUT-(A)dS metric while the Lorentzian {\em base metric} $ \mtrt$ has the same form but modified radial dependence.
In general, NUT parameters remain in both components.

We use this as an opportunity to  partially rehabilitate a `bad reputation' of NUT parameters which is based on the fact that in four dimensions they typically lead to causally ill-behaved spacetimes \cite{taub1951empty, newman1963empty, misner1963flatter}, see however recent discussion in \cite{Clement:2015cxa}. Similar pathology happens in higher dimensions when the NUT charges maintain their `Lorentzian character'. However,  when present only in the spatial part of the metric, the `Euclidean NUT' parameters just label deformations of the angular part of the metric and do not introduce regions with closed time-like curves. Such NUTs typically `smoothen' the curvature singularity, although the `axes of rotation' may suffer from non-regularity.

This is precisely what happens with the surviving NUT parameters after we take our limit. Those  surviving in the base part of the metric maintain their Lorentzian character and lead to standard pathologies. Those surviving in the seed part of the metric, which is Euclidean, play different geometric role: they do not lead to pathologies with closed time-like curves and tend to smoothen the curvature singularity. Such NUTs control deformation and also a {\em global twist} of the Euclidean geometry. Namely, in higher dimensions we have a freedom in making linear combinations and various periodic identifications of Killing coordinates, which we refer to as global twisting of the geometry. Since the metric contains non-diagonal terms, the freedom in twisting cannot be easily resolved. The presence of NUT parameters additionally causes non-regularity on the axes of rotation and thus enters non-trivially into the discussion of twisting.

Obviously, the twist is non-trivial only in the case when the seed metric is sufficiently `multi-dimensional', that is, when it possesses more than one azimuthal Killing coordinate. Since the number of Killing coordinates of the resulting seed metric equals (in the studied even-dimensional case) the number of rotation parameters that were {\em simultaneously} set equal to zero, to have a non-trivial twist two or more rotations have to be switched off simultaneously. We call the corresponding limit a {\em multiple-spin-zero limit}. Such a limit  leads to the {\em deformed and twisted} black holes. On the other hand, when the rotation parameters are switched off {\em successively} one by one, that is when we apply several {\em single-spin-zero limits}, the resulting seed metrics are two dimensional and their twist vanishes. In particular, we shall consider a case where all rotations and twists are eventually switched off. We call the corresponding solutions the {\em deformed} (non-twisted) black holes. Although they are static, their geometry is still non-trivially deformed by the presence of NUT charges. As we shall show in Sec.~\ref{sc:MultiLimit}, the same geometry can be obtained by a successive application of the multiple-spin-zero limit, each of which leaves the base metric only two dimensional.

After obtaining the limiting metrics we discuss their symmetries. In particular, we show that the resulting warped geometry admits a {\em degenerate} CCKY 2-form $\tens{\tilde h}$, inherited from the base metric $ \mtrt$. The resulting geometry therefore belongs to a wide class of {\em generalized Kerr-NUT-(A)dS spacetimes} constructed in \cite{HouriEtal:2008b,HouriEtal:2009} and further studied in \cite{Oota:2008uj}, see App.~A for their review. Despite the degeneracy of this tensor, we show that the new geometry inherits the full original symmetry structure of the Kerr-NUT-(A)dS family, namely, it possesses the full Killing tower of hidden and explicit symmetries that can be obtained through the scaling limits of the original Killing tower of symmetries. (These symmetries can be also built up by a proper combination of the Killing towers of both seed and base geometries,  see \cite{KrtousKubiznakKolar:2015}.) For this reason our new metrics represent a very special sub-class of generalized Kerr-NUT-(A)dS spacetimes, characterized by a property that they maintain the full Killing tower. In particular, this implies that the new class of solutions inherits the integrability properties of the original Kerr-NUT-(A)dS geometry.

As expected, for vanishing NUT, twist, and deformation parameters, the symmetry is further enlarged. For example, the Schwarzschild--Tangherlini-(A)dS solution  is spherically-symmetric and its group of isometries is larger than the symmetry group of the initial metric. Similar ``symmetry enlargement'' also happens, when one puts some smaller set of the rotation and NUT parameters equal to zero.

The plan of the paper is as follows. In the next section we introduce the Kerr-NUT-(A)dS geometry and analyze its basic properties. In Sec.~\ref{sc:Limitato0} the multiple-spin-zero limit (in which several rotation parameters are set equal to zero simultaneously) is exploited to obtain the new twisted and deformed black hole solutions. In Sec.~\ref{sc:MultiLimit} we discuss successive application of the multiple-spin-zero limits and arrive thus to the deformed black holes without twists and rotations. In Sec.~\ref{sc:D46} we present specific examples in 4 and 6 dimensions and Sec.~\ref{sc:Discussion} is devoted to discussion. In App.~\ref{apx:degPCCKYfam} we review the generalized Kerr-NUT-(A)dS spacetimes and show that the obtained metrics form their special subclass.  App.~\ref{apx:tech} contains additional technical results supporting the main text.

\section{An analysis of Kerr-NUT-(A)dS spacetimes}
\label{sc:KerrNutAdS}

\subsection{The metric}

In what follows, we concentrate on analyzing the even-dimensional Kerr-NUT-(A)dS spacetimes\footnote{%
A generalization to odd dimensions, where the Kerr-NUT-(A)dS metrics contain extra `odd' terms, is technically straightforward.}
with the spacetime dimensions parameterized as $D=2N$. The metric of such spacetime takes the following form \cite{ChenLuPope:2006}:\footnote{%
The Greek indices take values $\alpha,\,\beta,\,\mu,\,{\dots} =1,\dots, N$, the Latin indices from the middle of the alphabet run through $i,\,j,\,k,\,{\dots}= 0,\dots,N-1$. We do not use the Einstein summation convention for these indices, but we do not indicate limits in the sums and products explicitly. We do employ the Einstein summation convention for general spacetime indices ${a,\, b,\, c,\, \dots=1,\dots,2N}$ used later in the paper.}
\begin{equation}\label{KerrNUTAdSpsi}
    \mtrc = \sum_\mu\Biggl[\,\frac{\U_\mu}{X_\mu}\,\grad x_\mu^2+\frac{X_\mu}{\U_\mu}\biggl(\sum_k \A{k}_\mu\grad\psi_k\biggr)^{\!\!2}\,\Biggr]\;.
\end{equation}

A  variety of the {\em off-shell} metrics is hidden in the metric functions, $X_\mu$, each of which depends just on the corresponding 
coordinate $x_\mu$,
\be
{X_\mu=X_\mu(x_\mu)}\,.
\ee
The vacuum Einstein equations restrict these metric functions to be particular polynomials ${X_\mu=\mX{0}_\mu}$, where
\begin{equation}\label{mXdef}
    \mX{p}_\mu = \lambda \biggl[\Ja(x_\mu^2)-2\NUT_\mu a_\mu^2\,\Ua_\mu\, \Bigl(\frac{x_\mu}{a_\mu}\Bigr)^{1-2p}\biggr]\;.
\end{equation}
With this choice of $X_\mu$ we call the Kerr-NUT-(A)dS metric \eqref{KerrNUTAdSpsi} {\em on-shell}, highlighting the fact that it solves the Einstein equations. [Polynomials  $\mX{p}_\mu$ with general $p$ will emerge in the limiting process studied in the next section.]

The remaining metric functions $\A{k}_\mu$, $\U_\mu$ and auxiliary functions $\J_\mu(a^2)$ are explicit polynomials of coordinates $x_\mu$, defined by the following relations:\footnote{%
For brevity, we do not indicate dependence of $\J_\mu(a^2)$, $\A{k}_\mu$, and $\U_\mu$ on coordinates $x_\mu$. A completely explicit notation should be $\J_\mu(a^2;x_1,\dots,x_N)$, ${\A{k}_\mu(x_1,\dots,x_N)}$, etc. Similar remark also applies to functions
${\Ja_\mu(x^2)}$, 
${\Aa{k}_\mu}$, 
and ${\Ua_\mu}$ defined below.}
\begin{gather}\label{metricpolysmu}
    \J_\mu(a^2)=\prod_{\substack{\nu\\\nu\neq\mu}}(x_\nu^2-a^2)=\sum_k \A{k}_\mu (-a^2)^{N{-}1{-}k}\;, \\
    \U_\mu=\J_\mu(x_\mu^2)\;.
\end{gather}
We define also auxiliary functions $\J(a^2)$ and $\A{k}$,
\begin{equation}\label{metricpolys}
    \J(a^2)=\prod_{\nu}(x_\nu^2-a^2) =\sum_{k=0,\dots,N} \A{k} (-a^2)^{N{-}k}\;.
\end{equation}
Properties of these functions are listed in App.~\ref{apx:JAU}.

Inspecting the polynomials \eqref{mXdef}, we see that the on-shell Kerr-NUT-(A)dS metric
contains $N$ parameters $a_\mu$, these are related to {\em rotations}. They can be combined into polynomials $\Ja_\mu(x^2)$, $\Aa{k}_\mu$, $\Ua_\mu$, $\Ja(x^2)$, and $\Aa{k}$ defined in a manner analogous to functions $\J_\mu(a^2)$, $\A{k}_\mu$, $\U_\mu$, ${\J(a^2)}$, and ${\A{k}}$,  \eqref{metricpolysmu}--\eqref{metricpolys}, with $x_\mu$ and $a_\mu$ `interchanged'. For example,
\begin{gather}\label{parametricpolysmu}
    \Ja_\mu(x^2)=\prod_{\substack{\nu\\\nu\neq\mu}}(a_\nu^2-x^2)=\sum_k \Aa{k}_\mu (-x^2)^{N{-}1{-}k}\;, \\
    \Ua_\mu=\Ja_\mu(a_\mu^2)\;,
\end{gather}
 cf.\ App.~\ref{apx:JAU}.

The on-shell metric contains also the parameter~$\lambda$ related to the {\em cosmological constant} according to
\begin{equation}\label{Lambdalambda}
  \Lambda=(2N-1)(N-1)\lambda\;,
\end{equation}
and parameters~$\NUT_\mu$ which encode {\em mass} and what is usually vaguely called {\em NUT parameters} or {\em NUT charges}.

Let us note here, that we took advantage of the `vagueness' of NUT charges $\NUT_\mu$ and introduced a particular combination of parameters $\NUT_\mu\Ua_\mu a_\mu^{2p+1}$ in \eqref{mXdef} instead of writing just a simple parameter $b_\mu$, which is a common practice. The reason is that parameters $\NUT_\mu$  introduced in this way will have a trivial scaling behavior under the limit performed below whereas parameters $b_\mu$ would have to be nontrivially rescaled to obtain the desired result.

The metric \eqref{KerrNUTAdSpsi} can be also expressed using a different set of angular coordinates $\phi_\alpha$,
\begin{equation}\label{phipsirel}
    \phi_\alpha = \lambda a_\alpha \sum_k \Aa{k}_\alpha\psi_k\;,\quad
    \psi_k = \sum_\alpha\frac{(-a_\alpha^2)^{N{-}1{-}k}}{\Ua_\alpha}\frac{\phi_\alpha}{\lambda a_\alpha}\;.
\end{equation}
Since $\phi$'s are just constant linear combinations of $\psi$'s, they are also Killing coordinates. In the maximally symmetric case, they are related to $N$ independent planes of rotations. Using these angles, the metric takes the form
\begin{equation}\label{KerrNUTAdSphi}
    \mtrc = \sum_\mu\Biggl[\,\frac{\U_\mu}{X_\mu}\,\grad x_\mu^2+\frac{X_\mu}{\U_\mu}\biggl(\sum_\alpha \frac{\J_\mu(a_\alpha^2)}{\Ua_\alpha}\frac{1}{\lambda a_\alpha}\grad\phi_\alpha\biggr)^{\!\!2}\,\Biggr]\;.
\end{equation}

\subsection{The gauge freedom and counting of parameters}

Summarizing the previous subsection, the on-shell Kerr-NUT-(A)dS geometry is given by the metric \eqref{KerrNUTAdSpsi} or \eqref{KerrNUTAdSphi} with $X_\mu$ given by $\mX{0}_\mu$, \eqref{mXdef}. The geometry is labeled by parameters $a_\mu$ and $\NUT_\mu$. The remaining parameter $\lambda$ is fixed by the cosmological constant through \eqref{Lambdalambda}.

However, the parameters $a_\mu$ and $\NUT_\mu$ are not independent. There exist a one-parametric gauge freedom in rescaling coordinates, metric functions, and parameters which leaves the metric in the same form. Namely, it reads
\begin{gather}
   x_\mu \to s x_\mu\;,\quad \phi_\alpha\to\phi_\alpha\;,\quad\psi_k\to s^{-(k{+}1)}\psi_k\;,\notag\\
   a_\mu \to s a_\mu\;,\quad \NUT_\mu\to \NUT_\mu\;,\label{rescaling}\\
   X_\mu\to s^{2N} X_\mu\;,\quad\U_\mu\to s^{2(N{-}1)}\U_\mu\;,\quad\A{k}_\mu\to s^{2k}\A{k}\;.\notag
\end{gather}
One of the parameters $a_\mu$ can thus be set to a suitable value using this gauge freedom. We will use this freedom in the Lorentzian case, requiring gauge condition \eqref{aNnorm} below.
Thus, for a fixed cosmological constant the on-shell
Kerr-NUT-(A)dS  metrics form ($2N-1)$-parametric family of solutions of the Einstein equations.

Inspecting metric \eqref{KerrNUTAdSphi} with $X_\mu=\mX{0}_\mu$, we can also observe that the dependence on the cosmological parameter $\lambda$ is just a global scaling. Indeed, one can easily see that
\begin{equation}\label{globalscale}
  \lambda\mtrc = \mtrc|_{\lambda=1}\;.
\end{equation}
We will use this property mainly in the Euclidean case to eliminate a global scale in one part of the warped metric, see \eqref{mtrclimitsc}.

\subsection{The orthonormal frame and signature}

It is useful to introduce the following orthonormal frame of 1-forms:
\begin{gather}\label{1formfr}
    \enf{\mu} = \biggl(\frac{\U_\mu}{X_\mu}\biggr)^{\!\!\frac12}\grad x_\mu\;,\\
    \ehf{\mu} = \biggl(\frac{X_\mu}{\U_\mu}\biggr)^{\!\!\frac12}\sum_k\A{k}_\mu\grad \psi_k
              = \biggl(\frac{X_\mu}{\U_\mu}\biggr)^{\!\!\frac12}\sum_\alpha \frac{\J_\mu(a_\alpha^2)}{\Ua_\alpha}
                \frac{\grad{\phi_\alpha}}{\lambda a_\alpha}\;,\notag
\end{gather}
and the dual frame of vectors:
\begin{equation}\label{vecfr}
\begin{gathered}
    \env{\mu} = \biggl(\frac{X_\mu}{\U_\mu}\biggr)^{\!\!\frac12}\cv{x_\mu}\;,\\
\begin{aligned}
    \ehv{\mu} &= \biggl(\frac{\U_\mu}{X_\mu}\biggr)^{\!\!\frac12}\sum_k\frac{(-x_\mu^2)^{N{-}1{-}k}}{\U_\mu} \,\cv{\psi_k}\\
              &= \biggl(\frac{\U_\mu}{X_\mu}\biggr)^{\!\!\frac12}\sum_\alpha \frac{\Ja_\alpha(x_\mu^2)}{\U_\mu}
                 \,\lambda a_\alpha\,\cv{\phi_\alpha}\;.
\end{aligned}
\end{gathered}
\end{equation}
The form \eqref{KerrNUTAdSpsi} of the metric indicates that there is a natural splitting of the tangent space into 2-planes spanned on $\env{\mu}$ and $\ehv{\mu}$. These planes, which we call the {\em $\mu$-planes}, are characterized by antisymmetric 2-forms
\begin{equation}
    \plnf{\mu} = \enf{\mu}\wedge\ehf{\mu}\;.\label{plnf}
\end{equation}
The metric and the corresponding Levi-Civita tensor are then written as
\begin{equation}\label{mtrcLCT}
    \mtrc = \sum_\mu \Bigl(\enf{\mu}\enf{\mu}+\ehf{\mu}\ehf{\mu}\Bigl)\;,\quad \LCT = \plnf{1}\wedge\dots\wedge\plnf{N}\;.
\end{equation}

Although the expression \eqref{mtrcLCT} suggests that the (off-shell) Kerr-NUT-(A)dS metric is positive definite, it can actually describe geometries of various signatures. The normalization in \eqref{1formfr} and \eqref{vecfr} can be only `artificial' and the positively normalized vectors and 1-forms can be, in fact, imaginary. The real signature is given by signs of the metric functions $X_\mu/U_\mu$ and these depend on a particular choice of coordinate ranges. The signature can be also affected if we choose some of the coordinates to be Wick-rotated, i.e., purely imaginary.

We will be mainly interested in (i) geometries with the Euclidean signature and positive curvature and (ii) physical geometries of the Lorentzian signature and arbitrary sign of the curvature. Before specifying the coordinate ranges and Wick rotations in these cases, let us first identify the maximally symmetric geometry.

\subsection{Maximally symmetric spaces}

It has been already observed in \cite{HamamotoEtal:2007} that the class of Kerr-NUT-(A)dS geometries contains also a trivial case of the maximally symmetric spaces. It is obtained by setting the NUT and mass parameters equal to zero, $\NUT_\mu=0$, while the parameters $a_\mu$ remain unrestricted.
The metric functions $X_\mu$ then simplify into a common polynomial $\lambda\Ja(x^2)$ in the corresponding variable, ${X_\mu=\lambda\Ja(x_\mu^2)}$. The roots of this polynomial are exactly the parameters~$a_\mu^2$.

With this choice and employing the orthogonality relations \eqref{Jortrel3}, the metric \eqref{KerrNUTAdSphi} can be transformed into
\begin{equation}\label{KerrNUTAdSphiMS}
    \mtrc = \sum_\mu\Bigg[\,\frac{\U_\mu}{\lambda \Ja(x_\mu^2)}\,\grad x_\mu^2-\frac{J(a_\mu^2)}{\Ua_\mu}\frac{1}{\lambda a_\mu^2}\grad\phi_\mu^2\,\Biggr]\;.
\end{equation}
Introducing $N+1$ coordinates $\rho_\mu$, $\mu=0,\,1,\dots,\,N$, instead of $N$ coordinates $x_\mu$, and employing the Jacobi transformation\footnote{%
The expression for $\rho_0$ could be written in a way similar to other $\rho_\mu$, provided we extended the set of parameters $a_\mu$ by one additional parameter $a_0=0$.}
\begin{equation}\label{rhodef}
    \lambda\rho_\mu^2=\frac{J(a_\mu^2)}{-a_\mu^2\,\Ua_\mu}\;,\quad
    \lambda\rho_0^2 = \frac{J(0)}{\Ja(0)} = \frac{\A{N}}{\Aa{N}}\;,
\end{equation}
one can show that the new coordinates are restricted by the following constraint:
\begin{equation}\label{rhoconstraint}
    \lambda\sum_{\mu=0}^N\rho_\mu^2 = 1\,,
\end{equation}
while the metric takes a simple form
\begin{equation}\label{multicyl}
    \mtrc = \grad\rho_0^2 + \sum_{\mu} \Bigl[ \grad\rho_\mu^2 +  \rho_\mu^2\,\grad\phi_\mu^2\Bigr]\;.
\end{equation}
Clearly, $\rho_0$, $\rho_\mu$, $\phi_\mu$ are multi-cylindrical coordinates on  a $2N$-dimensional (pseudo-)sphere, given by the constraint~\eqref{rhoconstraint}, embedded into a $(2N{+}1)$-dimensional flat space.

\subsection{A Euclidean instanton}

Assuming that all coordinates $x_\mu$, $\psi_k$ are real and parameters $a_\mu$, $\NUT_\mu$, $\lambda$ are positive, the signature of the metric \eqref{KerrNUTAdSpsi} is determined by signs of the metric functions $X_\mu/U_\mu$. To determine these signs we need to specify ranges of coordinates $x_\mu$. All these coordinates, as well as corresponding parameters $a_\mu$ and $\NUT_\mu$, enter our metric in an equivalent way. However, it will be useful to relabel these quantities in such a way that parameters $a_\mu$ are ordered as follows,
\begin{equation}\label{aordeucl}
    0<a_1<a_2<\dots<a_N\;.
\end{equation}

In the maximally symmetric case $\NUT_\mu=0$, parameters $a_\mu$ coincide with the roots of the polynomials $X_\mu$, and determine axes as well as restrict the ranges of coordinates $x_\mu$. Zeros in functions $\U_\mu$ can be avoided be choosing
\begin{equation}\label{xrangesMS}
    -a_1<x_1<a_1<x_2<a_2<\dots<x_N<a_N\;.
\end{equation}
With angular coordinates restricted to a circle
\begin{equation}\label{phirangesMS}
    -\pi<\phi_\mu<\pi\;,
\end{equation}
the metric \eqref{KerrNUTAdSphi} describes a homogeneous metric on the Euclidean sphere with radius $\ell=1/\sqrt\lambda$, that is written in ``multi-elliptic'' coordinates parameterized by constants~$a_\mu$.

For sufficiently small positive values of parameters $\NUT_\mu$, the roots of metric functions $X_\mu$ change only slightly; they determine a more narrow range, now for each latitude coordinate $x_\mu$ different. Namely, we require
\begin{equation}\label{xrange}
    \xroot-\mu<x_\mu<\xroot+\mu\;,
\end{equation}
where $\xroot\pm\mu$ are roots of $X_\mu$ such that $a_{\mu-1}<\xroot-\mu<\xroot+\mu<a_\mu$ (with $\xroot-1<-a_1<\xroot+1<a_1$ for $\mu=1$). With this choice the metric \eqref{KerrNUTAdSpsi} represents the {\em Euclidean instanton} of signature $(++\dots+)$.

Parameters $\NUT_\mu$ encode deformations of the geometry, namely how it deviates from the geometry of the sphere. Parameters $a_\mu$ are for non-vanishing $\NUT_\mu$ also non-trivial---they do not label just a choice of coordinates, as in the maximally symmetric case, but they also change the geometry. For non-trivial values of $\NUT_\mu$ the geometry does not have a curvature singularity---the singularity of the Riemann tensor occurs outside of the coordinate range \eqref{xrangesMS} of the Euclidean instanton.

The global definition and regularity of the geometry described by metrics \eqref{KerrNUTAdSpsi} or \eqref{KerrNUTAdSphi} has to be concluded by specifying which Killing angles should be identified. Any linear combination of Killing coordinates  (with constant coefficients) forms again a Killing coordinate and it is not a~priory clear which of the Killing coordinates should be periodic. Typically, angles $\psi_k$ are \emph{not} those which should be periodic. Since Killing coordinates are non-trivially coupled in the metric (the metric is not diagonal in these directions), a particular choice of the periodicity of Killing coordinates can introduce a non-trivial twisting of the geometry, as well as possible irregularities on the axes.

We will not discuss these characteristics in more detail and for our purposes simply remember that the Euclidean instanton describes a deformed and twisted spherical-like geometry. (For more details we refer the interested reader to a vast literature on the subject of gravitational instantons, e.g. \cite{hawking1977gravitational, Page:1979zv, Page:1979aj, gibbons1979classification, eguchi1980gravitation, hunter1998action, Mann:1999pc, Chamblin:1998pz, Mann:2003zh, Mann:2005ra, Clarkson:2006zk, ChenLuPope:2007}.)

\subsection{Black hole solutions}\label{Blackholes}

The Lorentzian signature can be achieved by Wick-rotating some of the coordinates and parameters. Different choices can lead to different interpretations of the metric. We restrict ourselves to the case where the coordinate $x_N$ is Wick-rotated to a radial coordinate $r$ and the angular coordinate $\phi_N$ to the time coordinate~$t$. The remaining coordinates retain their original character. Namely, we set
\begin{equation}\label{Wick}
    x_N^2 = - r^2\;,\quad
    \phi_N = \lambda a_N t\;,
\end{equation}
with $r$ and $t$ real, and introduce a real mass parameter~$m$ by
\begin{equation}\label{mass}
    \NUT_N =m\lambda^{N{-}\frac32}\;.
\end{equation}
Notice that all coordinates $\psi_k$ remain real.
We also use the invariance  \eqref{rescaling} of the metric to correlate $a_N$ with~$\lambda$,
\begin{equation}\label{aNnorm}
    a_N^2 = -\frac1\lambda\;.
\end{equation}
Other parameters $a_\mu$ remain restricted as in the Euclidean case
\begin{equation}\label{aordlor}
    0<a_1<a_2<\dots<a_{N{-}1}<a_N\;,
\end{equation}
where the last inequality applies only for $a_N$ real (i.e., for $\lambda<0$). The ranges of coordinates $x_\mub$, $\mub=1,\dots, N-1$, are given again by roots of functions $X_\mub$, as in \eqref{xrange}.

With these choices metric \eqref{KerrNUTAdSphi} has the physical signature $(-+\dots+)$. More explicitly, it reads
\begin{widetext}
\begin{equation}\label{KerrNUTAdSphiWick}
\begin{split}
    \mtrc &=
    -\frac{\horfc}{\Sigma}\Biggl(\prod_\nub\frac{1+\lambda x_\nub^2}{1+\lambda a_\nub^2} \;\grad t
      - \sum_\nub \frac{\Jb(a_\nub^2)}{a_\nub(1+\lambda a_\nub^2)\Uab_\nub}\grad\phi_\nub\Biggr)^{\!\!2}
    +\frac{\Sigma}{\horfc}\,\grad r^2
    \\
    &\quad+\sum_\mub \frac{(r^2+x_\mub^2)\Ub_\mub}{-\mX{0}_\mub}\,\grad x_\mub^2
    +\sum_\mub\frac{-\mX{0}_\mub}{(r^2+x_\mub^2)\Ub_\mub}\Biggl(
      \frac{1-\lambda r^2}{1+\lambda x_\mub^2}\prod_\nub\frac{1+\lambda x_\nub^2}{1+\lambda a_\nub^2} \,\grad t
      + \sum_\nub \frac{(r^2+a_\nub^2)\Jb_\mub(a_\nub^2)}{a_\nub(1+\lambda a_\nub^2)\,\Uab_\nub} \grad\phi_\nub \Biggr)^{\!\!2}\;,
\end{split}
\end{equation}
\end{widetext}
with
\begin{gather}
    \horfc= -\mX{0}_N = \bigl(1{-}\lambda r^2\bigr)\prod_\nub\bigl(r^2{+}a_\nub^2\bigr)
       -2m\, r \prod_\nub\bigl(1{+}\lambda a_\nub^2\bigr) \;,\notag\\
    -\mX{0}_\mub = \bigl(1{+}\lambda x_\mub^2\bigr) \Jab(x_\mub^2)
       -2\NUT_\mub a_\mub  \bigl(1{+}\lambda a_\mub^2\bigr) \Uab_\mub\,x_\mub \;,\notag\\[1ex]
    \Sigma=\U_N=\prod_\nub(r^2+x_\nub^2)\;.\label{physmtrcfc}
\end{gather}
Here, the barred indices take values $\mub,\nub=1,\dots,N{-}1$ and functions $\Jb_\mub$, $\Ub_\mub$, $\Uab_\mub$ and $\Jab$ are defined by expressions \eqref{metricpolysmu} with restricted sets of coordinates $x_\mub$ and parameters $a_\mub$ (without $x_N$ and $a_N$).

\begin{figure}[b]
\includegraphics[width=3.3in]{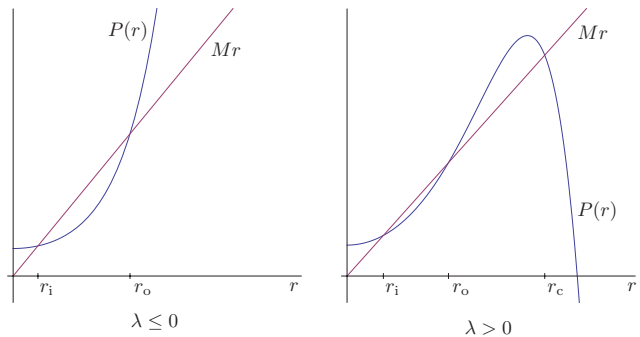}
\caption{\label{fig:horfc}%
{\bf Black hole horizons.}
The zeros of $\horfc$, \eqref{physmtrcfc}, determine horizons of the black hole. $\horfc$ has the following structure: $\horfc=P(r)-Mr$, where $P(r)$ is an even polynomial and $M$ is a constant. The polynomial $P(r)$ has no real roots for $\lambda\leq0$ and two real roots for $\lambda>0$. The zeros of $\horfc$ correspond to intersections of the polynomial and the straight halfline. For $\lambda\leq0$ there can be two intersections (outer and inner horizons), one touching intersection (extremal horizon) or no intersections (naked singularity). For $\lambda>0$ there is one additional intersection corresponding to the cosmological horizon.}
\end{figure}

The roots of the metric function $\horfc$ determine horizons. For $\lambda\leq0$ there can be two roots (outer and inner horizons), one double root (extremal horizon) or no roots (the case of naked singularity). For $\lambda>0$ there is always one additional root representing the cosmological horizon. See Fig.~\ref{fig:horfc}.

For $m\neq0$ and $\NUT_\mub=0$, $\mub=1,\dots,N-1$, the parameters $a_\mub$, $\mub=1,\dots,N-1$, can be identified with the rotation parameters of the black hole. However, for non-trivial NUT parameters $\NUT_\mub$ the interpretation is more unclear.  Similar to the Euclidean instanton, both sets of parameters $\NUT_\mub$ and $a_\mub$, as well as the specification of periodicity of Killing coordinates affect the deformation of the geometry. Since the Lorentzian signature is involved, the geometry has in general NUT-like irregular behavior including the existence of closed time-like curves around the axes.

\subsection{The Killing tower}
\label{sc:KillTower}

Let us conclude this section by a brief review of the symmetry structure of the (off-shell) Kerr-NUT-(A)dS spacetimes, see \cite{
KrtousEtal:2007a, Houri:2007uq, Krtous:2015,FrolovKubiznak:2008} for more details.
As mentioned in the introduction,
the {\em canonical form} of Kerr-NUT-(A)dS spacetimes \eqref{KerrNUTAdSpsi} is uniquely determined by
the existence of a non-degenerate CCKY 2-form  $\tens{h}$ \cite{KubiznakFrolov:2007, Houri:2007xz, KrtousFrolovKubiznak:2008}.\footnote{Roughly speaking, while coordinates $x_\mu$  are related to the eigenvalues of $\tens{h}$, coordinates $\psi_k$ are the Killing coordinates corresponding to the Killing vectors generated from $\tens{h}$ by formula \eqref{KVKTrel} below. The functional independence of all these coordinates is guaranteed by a `non-degeneracy' condition imposed on~$\tens{h}$.} This 2-form reads
\begin{equation}\label{PCCKY}
    \PCCKY = \sum_\mu x_\mu \plnf\mu
      = \sum_\mu x_\mu\grad x_\mu\wedge\sum_k \A{k}_\mu\grad\psi_k
      \;,
\end{equation}
and satisfies the CCKY equation
\begin{equation}\label{PCCKYeq}
    \nabla_a h_{bc} = g_{ab}\xi_{c} -g_{ac}\xi_{b}\;,
\end{equation}
with the {\em primary Killing--Yano 1-form $\PKV$}  given by 
\begin{equation}\label{PKVdef}
    \PKV = \frac{1}{D{-}1}\covd\cdot\PCCKY = \sum_\mu \frac{X_\mu}{\U_\mu}\sum_k \A{k}_\mu\grad\psi_k\;.
\end{equation}
Raising the index, $\PKV$ turns into a \emph{primary Killing vector}
\begin{equation}\label{PKVcoor}
    \PKV = \KV{0} = \cv{\psi_0}\;.
\end{equation}

The non-degenerate CCKY 2-form $\tens{h}$ generates a full {\em Killing tower} of symmetries \cite{KrtousEtal:2007a,Frolov:2008}. Namely, it defines
$N$ CCKY forms $\CCKY{k}$, ${k=0,\,\dots,\,N{-}1}$, of rank $2k$,
\begin{equation}\label{CCKYdef}
    \CCKY{k} = \frac1{k!} \frac1{\sqrt{\Aa{k}}}\,\PCCKY^{\wedge k}\;.
\end{equation}
Each of these CCKY forms then defines the following {Killing--Yano} (KY) forms $\KY{k}$ of rank $2(N-k)$:
\begin{equation}\label{KYdef}
    \KY{k} = \hodge\CCKY{k}\;,
\end{equation}
and the following rank-2 Killing tensors $\Kt{k}$:
\begin{equation}\label{KTdef}
    \Kt{k} = \KY{k}\pbullet{2N-2k-1}\KY{k}\;.
\end{equation}
Here $\pbullet{p}$ denotes the partial contraction of two antisymmetric forms in the first $p$ indices divided by $p!$,
\begin{equation}\label{partbullet}
  (\alpha\pbullet{p}\beta)_{a{\dots}b{\dots}}
    =\frac1{p!}\,\alpha_{n_1\dots n_pa\dots}\,\beta^{n_1\dots n_p}{}_{b\dots}\;,
\end{equation}
where the upper indices were raised using the metric. If no index $p$ is indicated, the full contraction in all indices is assumed. The primary Killing vector $\PKV$ and the above Killing tensors then define  $N$  Killing vectors $\KV{j}$ according to
\begin{equation}\label{KVKTrel}
    \KV{j} = \KT{j}\cdot\PKV\;.
\end{equation}
In particular, we find the following expressions for the Killing vectors and Killing tensors:
\ba
\KV{j}&=&\frac1{\Aa{j}}\,\cv{\psi_j}\;,\label{KVcoor}\\
\KT{j} &=& \frac1{\Aa{j}}\sum_\mu \A{j}_\mu \Bigl(\env{\mu}\env{\mu}+\ehv{\mu}\ehv{\mu}\Bigr) \label{KTcoor}\\
    \!&=&\!\sum_\mu \frac{\A{j}_\mu}{\Aa{j}} \Biggl[\,\frac{X_\mu}{\U_\mu}\,\cv{x_\mu}^{\;2}
       \!+\!\frac{\U_\mu}{X_\mu}\biggl(\sum_k \frac{(-x_\mu^2)^{N{-}1{-}k}}{\U_\mu}\cv{\psi_k}\biggr)^{\!\!2}\,\Biggr]\;.\nonumber
\ea
Note that, in expressions \eqref{KVcoor} and \eqref{KTcoor} we have a normalization prefactor $1/\Aa{j}$ that originates from a definition  \eqref{CCKYdef}. Such normalization is not used in the literature. Since this prefactor is just a constant (a combination of parameters $a_\mu$), it does not influence the Killing property of the defined objects. However, it will be useful in the limiting procedure below.

The above Killing tower of symmetries stands behind many of the remarkable integrability properties of Kerr-NUT-(A)dS spacetimes,
see \cite{FrolovEtal:2007,SergyeyevKrtous:2008,KolarKrtous:2015b, OotaYasui:2008a,Oota:2008uj,CarigliaEtal:2011a,CarigliaEtal:2011b, PageEtal:2007,KrtousEtal:2007a, Houri:2007uq, KrtousEtal:2007b,Krtous:2015} for more details.
In particular, the $N$ Killing vectors $\KV{j}$ together with the $N$ rank-2 Killing tensors $\KT{j}$ define $2N$ integrals for geodesic motion that are independent and in involution, making this motion completely integrable in the Liouville sense \cite{PageEtal:2007,KrtousEtal:2007a, Houri:2007uq, KrtousEtal:2007b,Krtous:2015}.

This property remains true for the twisted and deformed black holes obtained by a limiting procedure in the next section. Namely, as we shall see, all the objects from the Killing tower do survive the limit and correspondingly obey the same mutual commutation relations that guarantee the above integrability.

\section{Twisted and deformed black holes}
\label{sc:Limitato0}

\subsection{Multiple-spin-zero limit}

Let us now concentrate on a situation when the Kerr-NUT-(A)dS spacetimes describe the Lorentzian black hole solution, studied in Sec.~\ref{Blackholes}. In this case we have the following ordering of coordinates $x_\mu$ and parameters~$a_\mu$:
\begin{equation}\label{xrangesMSa}
    x_1<a_1<x_2<a_2<\dots<x_{N-1}<a_{N-1}\;.
\end{equation}
Parameters $a_\mu$ are closely related to rotations: for vanishing NUT parameters they are exactly rotations
\cite{GibbonsEtal:2004,GibbonsEtal:2005}, in the presence of NUTs a proper identification is more delicate.
Nevertheless, in both cases it seems that ``unspinning'' the rotations is equivalent to sending these parameters to zero.

However, it is obvious from \eqref{xrangesMSa} that our metrics \eqref{KerrNUTAdSpsi} or \eqref{KerrNUTAdSphiWick} are not suitable for a straightforward limit of vanishing parameters $a_\mu$, or for a limit when such parameters coincide. If such limits were performed naively, some of the coordinates $x_\mu$ would degenerate. Since $x_{\mu}$ are eigen-values of the principal CCKY tensor $\bf{h}$, the latter becomes degenerate in this limit. But what is worse, the metric itself would not be well defined. In the following we show that a well defined limit $a_\mu\to0$ can be achieved by a proper simultaneous scaling of corresponding coordinates $x_\mu$. We first perform the limit for the canonical metric \eqref{KerrNUTAdSpsi}, then specialize to the black hole case.

Namely, let us consider a {\em multiple-spin-zero} limit in which the first $\Nb$ parameters $a_\mub$ and the corresponding coordinates $x_\mub$ are homogeneously rescaled to zero, while the remaining $\Nt=N-\Nb$ parameters and coordinates are left unscaled but renamed to be indexed from $1$ to $\Nt$. The Killing coordinates $\phi_\mu$, NUT parameters $\NUT_\mu$, and $\lambda$ remain unscaled but they get also appropriately renamed.
We denote by ${\bar {\mathcal{M}}}$ a $2\Nb$-dimensional submanifold which emerges in the limiting procedure---it is spanned on the first $\Nb$ $\mu$-planes. Since we are sending the corresponding rotations to zero, we call it the \emph{unspinned sector}. The complementary $2\Nt$-dimensional submanifold $\tilde{\mathcal{M}}$ emerging as a space spanned on the unscaled $\mu$-planes will be called the \emph{regular sector}.
Similarly, we say that the scaled (unscaled) coordinates and parameters belong to the unspinned (regular) sector, respectively.

If $\zr$ is a scaling parameter which goes to zero, $\zr\to0$, the limit is characterized by
\begin{gather}
\notag
    \mspace{-30mu}\text{unspinned} \mspace{36mu}\text{regular}
\\
\label{parlimit}
\begin{aligned}
    a_\mub &= \zr\, \ab_\mu\;, &   a_{\Nb+\mut} & = \at_\mut\;,\\
    \NUT_\mub &= \NUTb_\mu\;, &   \NUT_{\Nb+\mut} & = \NUTt_\mut\;,
\end{aligned}
\\
\label{xphilimit}
\begin{aligned}
    x_\mub &= \zr\, \xb_\mu\;, &   x_{\Nb+\mut} & = \xt_\mut\;,\\
    \phi_\mub &= \phb_\mu\;, &   \phi_{\Nb+\mut} & = \pht_\mut\;.
\end{aligned}
\end{gather}
We use the barred indices $\mub,\nub,\dots$ and $\kb,\lb,\dots$ for the unspinned quantities, and the tilded indices $\mut,\nut,\dots$ and $\kt,\lt,\dots$ for regular quantities. They take the following values:
\begin{equation}\label{indices}
\begin{aligned}
    \mub,\nub,\dots&=1,\dots,\Nb\;, &   \mut,\nut,\dots&=1,\dots\Nt\;,\\
    \kb,\,\lb,\,\dots&=0,\dots,\Nb{-}1\;, &   \kt,\,\lt,\,\dots&=0,\dots,\Nt{-}1\;.
\end{aligned}
\end{equation}

Let $\approx$ denotes the equality valid up to the terms of the higher order in $\zr$. Employing relations \eqref{phipsirel} between angular coordinates $\phi_\mu$ and $\psi_k$, we find
\begin{equation}\label{psilimit}
    \psi_{\kt}\approx\pst_{\kt}\;,\quad
    \psi_{\Nt+\kb}\approx\zr^{-(2\kb+1)}\frac1{\Aat{\Nt}}\psb_{\kb}\;.
\end{equation}
Note that, comparing Eqs.\ \eqref{xphilimit} and \eqref{psilimit}, the new coordinates are `enumerated' differently for Latin and Greek indices. Namely, for Latin indices the unspinned sector corresponds to ``lower'' values and the regular sector to ``upper'' values of indices, whereas this is opposite for Greek indices.

At the same time, metric functions \eqref{metricpolysmu} and \eqref{metricpolys} give
\begin{gather}
\begin{aligned}\label{Ulimit}
    \U_\mub&\approx\zr^{2(\Nb{-}1)}\At{\Nt}\Ub_\mub\;,  &
    \U_{\Nb{+}\mut} &\approx (-\xt_\mut^2)^{\Nb} \Ut_\mut\;,
\end{aligned}\\[1ex]
\begin{aligned}\label{Alimit}
    \A{\kt}_\mub&\approx\At{\kt}\;,  &
    \A{\Nt{+}\kb}_\mub&\approx\zr^{2\kb}\At{\Nt}\Ab{\kb}_\mub\;, \\
    \A{\kt}_{\Nb{+}\mut}&\approx\At{\kt}_{\mut}\;,  &
    \A{\Nt{+}\kb}_{\Nb{+}\mut}&\approx\zr^{2(\kb{+}1)}\At{\Nt{-}1}_\mut\Ab{\kb{+}1}\;, \\
    \A{\kt}&\approx\At{\kt}\;,  &
    \A{\Nt{+}\kb}&\approx\zr^{2\kb}\At{\Nt}\Ab{\kb}\;,
\end{aligned}
\end{gather}
and analogous relations hold for $\Ua_\mu$, $\Aa{k}_\mu$, and $\Aa{k}$. Starting from \eqref{mXdef}, we also have
\begin{equation}\label{Xlimit}
    \mX{p}_\mub\approx\zr^{2\Nb}\At{\Nt}\mXb{p}_\mub\;,  \quad
    \mX{p}_{\Nb{+}\mut} \approx (-\xt_\mut^2)^{\Nb} \mXt{p+\Nb}_\mut\;.
\end{equation}
Hence, we have obtained the following relations for various subterms of the metric \eqref{KerrNUTAdSpsi}:
\begin{equation}\label{mtrctermslimit}
\begin{gathered}
    \frac{\mX{p}_\mub}{\U_\mub}\approx \zr^2\frac{\Aat{\Nt}}{\At{\Nt}}\frac{\mXb{p}_\mub}{\Ub_\mub}\;,\qquad
    \frac{\mX{p}_{\Nb{+}\mut}}{\U_{\Nb{+}\mut}}\approx \frac{\mXt{p{+}\Nb}_\mut}{\Ut_\mut}\;,\\
    \sum_k\A{k}_\mub\grad\psi_k\approx\frac1\zr\frac{\At{\Nt}}{\Aat{\Nt}}\sum_\kb\Ab{\kb}_\mub\grad\psb_\kb\;,\\
    \sum_k\A{k}_{\Nb{+}\mut}\grad\psi_k\approx\sum_\kt\At{\kt}_\mut\grad\pst_\kt\;.
\end{gathered}
\end{equation}
In all these expressions barred (or tilded) functions are defined in terms of $\ab_\mub$ and $\xb_\mub$ (or $\at_\mut$ and $\xt_\mut$) in a way analogous to how unbarred functions are defined in terms of $a_\mu$ and $x_\mu$. In other words, in both regular and unspinned sectors we use the same functional definitions, using the appropriate coordinates and parameters for a given sector.

\subsection{Twisted and deformed black holes}
Putting everything together, we find that in the leading order of $\sigma$ expansion the on-shell Kerr-NUT-(A)dS metric decouples to a warped  product of the base and seed metrics in the regular  and unspinned sectors,
\begin{equation}\label{mtrclimit}
    \mtrc \approx \mtrt + \warp^2 \mtrb\;.
\end{equation}
In the regular sector, the base metric $\mtrt$ is the off-shell $2\Nt$-dimensional Kerr-NUT-(A)dS metric spanned on coordinates $\xt_\mut$, $\pst_\kt$, characterized by parameters $\at_\mut$, $\NUTt_\mut$, and metric functions
\begin{equation}\label{Xtilded}
    \Xt_\mut=\mXt{\Nt}_\mut\;.
\end{equation}
In the unspinned sector, the seed metric $\mtrb$ is the on-shell $2\Nb$-dimensional Kerr-NUT-(A)dS metric spanned on $\xb_\mub$, $\psb_\kb$, characterized by  parameters $\ab_\mub$, $\NUTb_\mub$, and metric functions
\begin{equation}\label{Xbarred}
    \Xb_\mub=\mXb{0}_\mub\;.
\end{equation}
The warp-factor $\warp$ is
\begin{equation}\label{warpdef}
    \warp^2 = \frac{\At{\Nt}}{\Aat{\Nt}}=\frac{\xt_1^2\dots\xt_\Nt^2}{\at_1^2\dots\at_\Nt^2}\;.
\end{equation}
It depends only on the tilded variables, i.e., on regular sector.
The Levi-Civita tensor associated with the obtained metric reads
\begin{equation}\label{LCTlimit}
    \LCT\approx\warp^{2\Nb} \,\LCTt\wedge\LCTb\;.
\end{equation}

One can easily show that the obtained warped geometry \eqref{mtrclimit} still satisfies the vacuum Einstein equations with the cosmological constant given by \eqref{Lambdalambda}, for more details see \cite{KrtousKubiznakKolar:2015}. One can also calculate the Riemann tensor and the Kretschmann scalar for such metrics, as described in App.~\ref{KretschSec}.

For the Lorentzian signature, it is useful to eliminate the dependence of the unspinned metric $\mtrb$ on the curvature scale $\lambda$. After the Wick rotation \eqref{Wick} and normalization \eqref{aNnorm}, the warp-factor $\warp$ becomes
\begin{equation}\label{warpr}
    \warp^2 = \lambda\, {\tilde{r}}^2\, \frac{\xt_1^2\dots\xt_{\Nt{-}1}^2}{\at_1^2\dots\at_{\Nt{-}1}^2}\;.
\end{equation}
As we discussed in \eqref{globalscale}, metric $\lambda\mtrb=\mtrb|_{\lambda=1}$ is independent of $\lambda$. Thus,
\begin{equation}\label{mtrclimitsc}
    \mtrc \approx \mtrt + {\tilde{r}}^2\, \frac{\xt_1^2\dots\xt_{\Nt{-}1}^2}{\at_1^2\dots\at_{\Nt{-}1}^2}\; \mtrb|_{\lambda=1}\;.
\end{equation}
The unspinned metric $\mtrb$ describes the on-shell Euclidean instanton, i.e., a deformed twisted sphere. In the black hole case, the regular metric $\mtrt$ has Lorentzian signature and describes a part of the metric which includes time coordinate, radial coordinate, and rotating directions. Notice, that the regular metric belongs only to the off-shell class of Kerr-NUT-(A)dS metrics since the metric functions $\Xt_\mu$ are modified and their form depends on the dimensionality of the unspinned sector, cf.~\eqref{Xtilded}.

Inspecting metric \eqref{mtrclimit}, we observe that although we took the limit of vanishing $a_\mub=\zr\ab_\mub\to0$, the unspinned parameters $\ab_\mub$ remain in the resulting metric. The limit only ``decouples'' both sets of regular and unspinned variables and parameters.
Counting parameters in the obtained metric, we have $2\Nt$ parameters $\at_\mut$, $\NUTt_\mut$ labeling regular metric $\mtrt$ and $2\Nb$ parameters $\ab_\mub$, $\NUTb_\mub$ labeling unspinned metric $\mtrb$. However, now we can use rescaling freedom \eqref{rescaling} for both parts of the metric independently. Two of the parameters are therefore superfluous and they can be fixed by gauge fixing conditions. The limiting procedure therefore  eliminated only one parameter from the original metric.

We have thus obtained a new rich class of higher-dimensional black hole solutions that solve the vacuum  Einstein equations with the cosmological constant and contain one parameter less than the original Kerr-NUT-(A)dS spacetimes.
Because of the presence of NUT parameters the new spacetimes represent a non-trivial generalization of Kerr-AdS black holes obtained in \cite{MyersPerry:1986, GibbonsEtal:2004, GibbonsEtal:2005} which no longer belongs, due to the singular character of the scaling limit, to the Kerr-NUT-(A)dS class.

\subsection{Non-rotating black holes with twist}

An important special case of the limit occurs for $\boldsymbol{\Nt=1}$, $\boldsymbol{\Nb=N-1}$. Since we are interested in the physical signature, we employ the normalization \eqref{aNnorm}, and the Wick rotation \eqref{Wick},
\begin{equation}\label{Nt=1coors}
\begin{gathered}
    \lambda r^2 = \frac{x_N^2}{a_N^2}=\frac{\xt_1^2}{\at_1^2}=\warp^2\;,\\
    t = a_N \phi_N= a_N \pht_1 \approx \psi_0 = \pst_0 \;.
\end{gathered}
\end{equation}
Introducing the mass parameter \eqref{mass}, the metric function ${\Xt_1=\mXt{\Nb}_1 = -\horfcd}$ can be rewritten as
\begin{equation}\label{horfcNt=1}
    \horfcd =  1 - \lambda r^2 - \frac{2 m}{r^{2N-3}}\;.
\end{equation}
With these definitions, the limiting metric takes the following form:
\begin{equation}\label{mtrclimitNt=1}
    \mtrc\approx -\horfcd\grad t^2 + \frac1\horfcd \grad r^2
       + r^2 \mtrb|_{\lambda=1}\,.
\end{equation}
It describes a deformed and twisted static black hole with angular part given by the Euclidean instanton $\mtrb|_{\lambda=1}$.

In particular for $\NUTb_\mub=0$, the unspinned metric simplifies to a maximally symmetric geometry, that is, $\mtrb|_{\lambda=1}$ describes the Euclidean \mbox{$2\Nb$-dimensional} sphere of unit radius. We thus recovered the Schwarzschild--Tangherlini solution describing a \mbox{non-rotating} higher-dimensional black hole. The metric $\mtrb$ on the sphere is, however, written in multi-eliptical coordinates that are characterized by constants $\ab_\mub$.

\subsection{Limit of the Killing tower}\label{LimitKTower}

We now turn to the limit of the symmetry objects in the Killing tower. Obviously, since in the limit $\sigma\to 0$ the first $\Nb$ eigenvalues $x_{\mub}$ of the CCKY tensor $\PCCKY$ vanish, this tensor becomes a degenerate
CCKY tensor $\PCCKYt$ and the resulting metric has to belong to the class of generalized Kerr-NUT-(A)dS spacetimes \cite{HouriEtal:2008b,HouriEtal:2009}, reviewed in App.~\ref{apx:degPCCKYfam}. However, as we show now, the obtained metric is a very distinguished one in this class as it possesses an enhanced symmetry of the full Killing tower. In particular, we show that the limiting  $\PCCKYt$ can be understood as a non-degenerate CCKY on the submanifold~$\tilde{\mathcal{M}}$ and that another 2-form $\tens{\bar h}$ emerges in the limit and effectively plays the role of a non-degenerate CCKY tensor on the unspinned submanifold $\bar{\mathcal{M}}$. (See App.~\ref{apx:KillTowLim} for more details.)

First, employing \eqref{xphilimit} and \eqref{mtrctermslimit}, we obtain the following expressions for 2-forms $\plnf\mu$:
\begin{equation}\label{plnflimit}
    \plnf\mub \approx \warp^2 \, \plnfb\mub\;,\quad
    \plnf{\Nb+\mut} \approx \plnft\mut\;.
\end{equation}
The unspinned 2-forms $\plnfb\mub$ are normalized in the sense of metric $\mtrb$. The warp factor guarantees the normalization of $\plnf\mub$ with respect to $\mtrc$.

Starting from the expression \eqref{PCCKY} for $\PCCKY$, only terms with regular coordinates $x_{\Nt{+}\mut}=\xt_\mut$ survive in the leading order and we get
\begin{equation}\label{PCCKYlimit}
    \PCCKY\approx \PCCKYt = \sum_\mut \xt_\mut\, \plnft\mut \;.
\end{equation}
The limit of the primary Killing-Yano form $\PKV$ is straightforward, cf.~\eqref{PKVdef} and \eqref{mtrctermslimit},
\begin{equation}\label{PKVlimit}
    \PKV\approx \tilde\PKV \;.
\end{equation}
The limit \eqref{PCCKYlimit} of the CCKY form $\PCCKY$ is degenerate: it vanishes in the unspinned sector. Therefore the limiting metric does not have the canonical form \eqref{KerrNUTAdSpsi} and the obtained $\PCCKYt$ cannot be used for a direct construction of the Killing tower. [This tensor can, however, be understood as a non-degenerate CCKY 2-form on the base manifold $\tilde{\mathcal{M}}$, c.f. \eqref{PCCKYlimit}]. Nevertheless, we can find the limits of all objects $\CCKY{k}$, $\KY{k}$, $\KT{k}$, and $\KV{k}$ directly.

Had we used the standard definitions of the Killing tower objects without a prefactor given by suitable powers of $\Aa{k}$ [see the discussion below  expressions \eqref{KVcoor} and \eqref{KTcoor}], we would find that some of the objects in the limit vanished. However, it would be possible to extract the coefficient of the leading term in the $\zr$ expansion. Instead, this can be simply achieved by including the discussed prefactors into
the definitions of Killing objects---under the limit these prefactors supply necessary powers of the scaling parameter $\sigma$ to obtain a finite result.

After the limit, the Killing tower naturally splits into two sets, one associated with the metric $\mtrb$ of the unspinned sector and the other associated with the regular metric $\mtrt$. Here, we present only the resulting expressions, details of the limiting process can be found in App.~\ref{apx:KillTowLim}. The limits are
\begin{equation}\label{CCKYlimit}
\begin{aligned}
   \CCKY{\kt}       &\approx    \CCKYt{\kt}\;,\\
   \CCKY{\Nt{+}\kb} &\approx    \warp^{2\kb+1}\,\LCTt\wedge \CCKYb{\kb}\;.
\end{aligned}
\end{equation}
Similarly, for the KY forms we obtain
\begin{equation}\label{KYlimit}
\begin{aligned}
   \KY{\kt}          &\approx    \warp^{2\Nb}\,  \KYt{\kt}\wedge\LCTb\;,\\
   \KY{\Nt{+}\kb}    &\approx    \warp^{2(\Nt{-}\kb){+}1}\, \KYb{\kb}\;.
\end{aligned}
\end{equation}
The Killing tensors are quadratic in KY forms according to \eqref{KTdef}. The limit gives
\begin{equation}\label{KTlimit}
\begin{aligned}
   \KT{\kt}         &\approx    \KTt{\kt}
                 + {\textstyle\frac{\At{\kt}}{\Aat{\kt}}}\,\warp^{-2}\, \mtrb^{-1}\;,\\
   \KT{\Nt{+}\kb}   &\approx   \KTb{\kb}\;.
\end{aligned}
\end{equation}
Finally, we perform the limit of Killing vectors. In this case, however, even after employing the normalization factor $\Aa{k}$ in \eqref{KVcoor}, the limit of Killing vectors related to the unspinned coordinates is not finite\footnote{%
One could have included an additional prefactor for $\KV{\Nt+\kb}$ which would behave as $\zr$. However, we do not see any natural simple candidate for this, so we leave the explicit ${\zr}$-dependence in final expressions.}
\begin{equation}\label{KVlimit}
\begin{aligned}
   \KV{\kt}        &\approx \KVt{\kt} = {\textstyle\frac1{\Aat{\kt}}} \cv{\pst_\kt}\;,\\
   \KV{\Nt{+}\kb}  &\approx   \zr\, \KVb{\kb} = \zr\, {\textstyle\frac1{\Aab{\kb}}} \cv{\psb_\kb}\;.
\end{aligned}
\end{equation}
Nevertheless, the leading term of the Killing vectors $\KV{\Nt+\kb}$ is proportional to the Killing vector $\KVb{\kb}$ of the unspinned part $\mtrb$ of the metric. The limiting metric thus admits at least the same number of explicit Killing vectors as the original one, only their relation to the original Killing vectors is slightly more complicated.

It is possible to check that both expressions \eqref{KYdef} and \eqref{KTdef}, relating CCKY forms ${\CCKY{j}}$, KY forms ${\KY{j}}$, and Killing tensors ${\KT{j}}$ remain valid after the limit. The relation \eqref{KVKTrel} between the Killing vectors ${\KV{j}}$ and Killing tensors ${\KT{j}}$ holds also in the leading order. However, with the normalization \eqref{KVlimit} it remains non-degenerate only for vectors related to regular coordinates, i.e., for ${j=\jt=0,\dots,\Nt-1}$. For ${j=\Nt+\jb=\Nt,\dots,N-1}$ this relation degenerates in the zeroth order and one has to study next order terms in ${\PKV}$ and ${\KT{\Nt{+}\kb}}$ to establish its validity. 

To summarize, we have found that although the limiting metric $\tens{g}$ does not admit a \emph{non-degenerate} CCKY form, it possesses the full Killing tower of symmetries that is similar to the Killing tower of the original metric. It splits into two sectors which are closely related to the Killing towers of regular metric $\mtrt$ and unspinned metric $\mtrb$, see relations \eqref{CCKYlimit}--\eqref{KVlimit}.
Since the resulting Killing tower is obtained by the limiting procedure, the surviving symmetries satisfy the same mutual relations as the original symmetries, namely, they mutually commute. This then guarantees integrability and separability properties similar to the those of the original
Kerr-NUT-(A)dS spacetimes \cite{FrolovEtal:2007,SergyeyevKrtous:2008,KolarKrtous:2015b, OotaYasui:2008a,Oota:2008uj,CarigliaEtal:2011a,CarigliaEtal:2011b, PageEtal:2007,KrtousEtal:2007a, Houri:2007uq, KrtousEtal:2007b,Krtous:2015,FrolovKubiznak:2008}.
The integrability and separability properties based on the Killing tower of a warped space \eqref{mtrclimit} has been also recently discussed in a slightly broader context of metrics that admit complete separability of the Klein--Gordon equation \cite{KolarKrtous:2015b}.

\section{Deformed black holes}
\label{sc:MultiLimit}

Our aim in this section is to completely switch off all the rotations and twists, constructing so the static deformed non-twisted black holes whose deformations are caused by the presence of non-trivial NUT parameters.

In the previous section, we have studied the multiple-spin-zero limit in which several unspinned coordinates $x_\mu$ and their corresponding parameters were simultaneously rescaled to zero. We have found that the unspinned parameters survived this limit, but they decoupled from the regular coordinates and parameters.
It would be natural to study now a limit in which we switch off just one rotation but perform such a limit successively several times, finally switching-off all the rotations. Since we have seen that NUT parameters survive the limit, the resulting metric should contain remnants of them. However, to see details of such a procedure we would need to generalize the limit discussed in the previous section to the case of general metric polynomial ${\mX{p}_\mu}$, cf.\ Eq.~\eqref{mXdef}. Indeed, already after the first ${\Nb=1}$ limit, the base metric ${\mtrt}$ on ${2\Nt=2N-2}$ dimensional manifold ${\tilde{\mathcal{M}}}$ is an off-shell metric with metric functions given by ${\mX{1}_\mu}$. To perform the next ${\Nb=1}$ limit we would need to generalize the limit from the previous section to this more general case (we only studied the on-shell case with ${\mX{0}_\mu}$).

Such a generalization is possible and the successive application of the single-spin-zero limit can be performed. Interestingly, it turns out that the resulting geometry can be obtained also by a different limiting procedure in which we repeatedly rescale all coordinates except one. That is, we repeatedly perform the multiple-spin-zero limit ${\Nb=N-1}$, $\Nt=1$: in the first step applied to the original metric, in the second step to the unspinned metric $\mtrb$, etc. In the following we will present details of such a limit.

The metric obtained by this combined limiting process will be denoted by $\mtro$. In the Lorentzian case, it describes the deformed (but non-twisted) black holes. Let $\CCKYo{j}$, $\KYo{j}$, and $\KTo{j}$ are limits of CCKY forms $\CCKY{j}$, 
KY forms $\KY{j}$, 
and Killing tensors $\KT{j}$, 
respectively. They turn out to be finite and they are CCKY forms, KY forms, and Killing tensors of the limiting metric $\mtro$.

\subsection{Atomic 2-metrics}

To write down explicitly the limiting metric $\mtro$ and its symmetries, we first introduce some auxiliary objects. Let ${}^\alpha\mtrc$ denotes the  canonical Kerr-NUT-(A)dS metric \eqref{KerrNUTAdSphi} in $2\alpha$ number of dimensions, spanned on coordinates $x_1,\dots,x_\alpha$ and $\phi_1,\dots,\phi_\alpha$. Let ${}^\alpha\LCT$ represents the corresponding Levi-Civita tensor and ${}^\alpha\CCKY{j}$, ${}^\alpha\KY{j}$, ${}^\alpha\KT{j}$, $j=0,\dots,\alpha$ the symmetry objects in the corresponding Killing tower.
We introduce `auxiliary' $2\alpha$-dimensional metrics ${}^\alpha\mtro$ and ${}^\alpha\mtrd$ and the corresponding Levi-Civita tensors ${}^\alpha\LCTo$ and ${}^\alpha\LCTd$ by the following expressions:
\begin{gather}
\begin{aligned}
  {}^\alpha\mtro &= \mtrq_\alpha + \sx_\alpha^2 \mtrq_{\alpha{-}1} + \sx_\alpha^2 \sx_{\alpha{-}1}^2 \mtrq_{\alpha{-}2}
     +\dots +\sx_\alpha^2\dots\sx_2^2\mtrq_1\;,\\
  {}^\alpha\LCTo &= \LCTq_\alpha \wedge \sx_\alpha^2 \LCTq_{\alpha{-}1} \wedge \sx_\alpha^2 \sx_{\alpha{-}1}^2 \LCTq_{\alpha{-}2}
     \wedge\dots\wedge \sx_\alpha^2\dots\sx_2^2\LCTq_1\;,
\end{aligned}\label{omtrLCT}\\
\begin{aligned}
  {}^\alpha\mtrd &= \mtrq_N + \sx_N^2 \mtrq_{N{-}1} + \sx_{N}^2 \sx_{N{-}1}^2 \mtrq_{N{-}2}+\dots \\
     &\mspace{133.5mu}+\sx_N^2\dots\sx_{N{-}\alpha{+}2}^2\mtrq_{N{-}\alpha{+}1}\;,\\
  {}^\alpha\LCTd &= \mtrq_N + \sx_N^2 \LCTq_{N{-}1} \wedge \sx_{N}^2 \sx_{N{-}1}^2 \LCTq_{N{-}2}\wedge\dots \\
     &\mspace{132mu}\wedge\sx_N^2\dots\sx_{N{-}\alpha{+}2}^2\LCTq_{N{-}\alpha{+}1}\;.
\end{aligned}\label{dmtrLCT}
\end{gather}
The above definitions are expressed in terms of rescaled coordinates ${\sx_\mu=\frac{x_\mu}{a_\mu}}$ and $\phi_\mu$, through {\em atomic 2-metrics} $\mtrq_\mu$ and \mbox{2-forms}~$\LCTq_\mu$
\begin{equation}\label{qmtrLCTdef}
\begin{gathered}
  \mtrq_\mu = \frac1\lambda\Bigl(\frac1\Delta_\mu\grad\sx_\mu^2+\Delta_\mu \grad\phi_\mu^2\Bigr)\;,\\
  \LCTq_\mu = \frac1\lambda \grad\sx_\mu\wedge \grad\phi_\mu\;,
\end{gathered}
\end{equation}
where
\begin{equation}\label{Deltadef}
    \Delta_\mu = 1 - \sx_\mu^2 - 2 \NUT_\mu \sx_\mu^{-2\mu+3}\;.
\end{equation}

\subsection{Deformed black holes}

Now we summarize the results of the combined limiting process, the derivation is delegated to App.~\ref{apx:multilimit}.
The limiting metric is  $\mtro={}^N\mtro={}^N\mtrd$. The explicit expressions for it and the Levi-Civita tensors are
\begin{equation}\label{limmtrc}
\begin{aligned}
  \mtro &= \mtrq_N + \sx_N^2\, \mtrq_{N{-}1} +\dots +\sx_N^2\dots\sx_2^2\,\mtrq_1\;,\\
  \LCTo &= \LCTq_N \wedge \sx_N^2\, \LCTq_{N{-}1} \wedge \dots\wedge \sx_N^2\dots\sx_2^2\,\LCTq_1\;.
\end{aligned}
\end{equation}
If we perform the Wick rotation \eqref{Wick}, and assume normalization \eqref{aNnorm}, we can write the limit of the Lorentzian metric as
\begin{equation}\label{STmtrcmultilim}
\begin{split}
    &\mtro = -\horfcd\grad t^2+\frac1\horfcd\grad r^2\\
       &+ r^2 \Bigl[\mtrq_{N{-}1}
       +\sx_{N{-}1}^2\Bigl( \mtrq_{N{-}2} + \dots
       +\sx_3^2 (\mtrq_2+\sx_2^2 \mtrq_1)\Bigr)\Bigr]\Big\vert_{\lambda=1}\,,
\end{split}\raisetag{11ex}
\end{equation}
with $\horfcd$ given by \eqref{horfcNt=1}.
Here, coordinates $\xi_\mu$, $\mu>1$, run between roots of $\Delta_\mu$, \eqref{Deltadef}, which are for small positive $\NUT_\mu$ inside interval $(0,1)$. This restriction does not apply to ${\xi_N}$ in the Lorentzian case, since in that case ${\xi_N}$ is Wick rotated to radial coordinate ${r}$. The range of $\xi_1$ is inside $(-1,1)$ and the metric $\mtrq_1$ is trivial, see the discussion of metric $\mtrb$ in section \ref{ssc:D4}.

The curvature of $\mtro$ can be obtained by repetitive application of the warped splitting while employing relations \eqref{eq:warpRiem}--\eqref{eq:warpKret}, to get the following expression for the Riemann tensor:
\begin{equation}\label{RiemMlim}
\begin{split}
  \Riem
    &= -\frac\lambda4 \sum_{\mu=1}^N\Delta_\mu''
        \biggl(\prod_{\nu=\mu+1}^N \xi_\nu^2\biggr)
        \mtrq_\mu {\wedge}\!\!\!{\wedge}\, \mtrq_\mu\\
      &-\frac\lambda2\sum_{\mu=2}^N \frac{\Delta_\mu'}{\xi_\mu}\sum_{\nu=1}^{\mu-1}
        \biggl(\prod_{\alpha=\nu+1}^N \xi_\alpha^2\biggr)
        \mtrq_\mu {\wedge}\!\!\!{\wedge}\, \mtrq_\nu\\
      &-\frac\lambda2\sum_{\mu=2}^N \Delta_\mu
        \sum_{\alpha,\beta=1}^{\mu-1}
        \biggl(\prod_{\kappa=\alpha+1}^N \xi_\kappa^2\biggr)
        \biggl(\prod_{\lambda=\beta+1}^{\mu-1} \xi_\lambda^2\biggr)
        \mtrq_\alpha {\wedge}\!\!\!{\wedge}\, \mtrq_\beta
      \;,
\end{split}\raisetag{14ex}
\end{equation}
where
for two 2-forms ${\tens{p}}$, ${\tens{q}}$ we defined
\be
{(\tens{p}\,{\wedge}\!\!\!{\wedge}\,\tens{q})_{abcd}}={4\,\tens{p}_{[a|[c}\tens{q}_{d]|b]}}\,.
\ee
Substituting the expressions for $\Delta_\mu$, \eqref{Deltadef}, we can analyze the singular behavior of the curvature. In our case, it is sufficiently represented by the singularities of the Kretschmann scalar, which simplifies to
\begin{equation}\label{KretschmannMlim}
    \mathcal{K} = 4N(2N{-}1)\lambda^2+\lambda^2\sum_{\mu=2}^N
    \frac{16(\mu{-}1)^2(2\mu{-}1)(2\mu{-}3)\beta_\mu^2}
    {\xi_\mu^{4\mu-2}\bigl(\prod_{\alpha=\mu+1}^N\xi_\alpha^2\bigr)^2}\;.
\end{equation}

For non-trivial parameters $\NUT_\mu$ we thus obtained a deformed geometry which is, however, not twisted as the angular Killing coordinates $\phi_\mu$ do not mix among themselves (the metric is diagonal). For the Lorentzian signature the geometry has a curvature singularity at ${r=0}$. Inspecting the expression for the Kretschmann scalar \eqref{KretschmannMlim}, we see that the singularity occurs for $\xi_\mu=0$, $\mu>1$, which is out of the coordinate ranges, except for the radial coordinate ${r}$.

For vanishing NUT parameters, ${\NUT_\mu=0}$, ${\mu=1,\dots,N{-}1}$, one can introduce latitude angles~$\theta_\mu$,
\begin{equation}\label{thtdef}
    \sx_\mu = \cos\theta_\mu
\end{equation}
and the atomic 2-metric \eqref{qmtrLCTdef} simplifies to the homogeneous metric on a 2-sphere
\begin{equation}\label{2sphere}
    \mtrq_\mu = \grad\theta_\mu^2+\sin^2\!\theta_\mu \,\grad\phi_\mu^2\;.
\end{equation}

The limits of the CCKY forms, KY forms and Killing tensors are
\begin{align}
    \CCKYo{j} &= {}^j w \;\, {}^j\LCTd \;,\label{CCKYmultilim}\\
    \KYo{j} &= {}^j w^{2(N{-}j){+}1}\;\,{}^{N{-}j}\LCTo\;,\label{KYmultilim}\\
    \KTo{j} &= {}^{N{-}j}\mtro^{-1}\;.\label{KTmultilim}
\end{align}
They satisfy relations \eqref{KYdef} and \eqref{KTdef}.
Relation \eqref{KVKTrel} between Killing vectors and Killing tensors degenerates since in the limit Killing vectors $\KV{j}$ vanish except for $j=0$. However, we can identify immediately another set of $N$ explicit Killing vectors of the limiting metric $\mtro$, namely vectors $\cv{\phi_\mu}$.

\section{Explicit examples}
\label{sc:D46}
After presenting our general limiting procedure, it is helpful to demonstrate it on a couple of simple examples. We first apply the formalism in physical dimension ${D=4}$, where we just recover the standard solutions. In the next subsection we study the ${D=6}$ case, where we can identify more interesting alternatives.

\subsection{${D=4}$ dimensions}
\label{ssc:D4}
Our starting point is the 4-dimensional on-shell Kerr-NUT-(A)dS metric with Lorentzian signature ${(-+++)}$. Since in lower dimensions it is more natural to label coordinates by letters instead of indexing, we use the following notation: ${x_1 \to x}$, ${a_1\to a}$, $\psi_0\to t, \psi_1\to \psi$. The metric \eqref{KerrNUTAdSphiWick} then reads
\begin{equation}\label{KerrNUTAdSN2}
\begin{split}
  \mtrc =
    &-\frac{\horfc}{\Sigma}\bigl(\grad t + x^2\grad\psi\bigr)^2 + \frac\Sigma\horfc\,\grad r^2\\
    &+\frac\Sigma {\cal S}\,\grad x^2+\frac{{\cal S}}{\Sigma}\bigl(\grad t - r^2 \grad\psi\bigr)^2\;,
\end{split}
\end{equation}
where
\begin{equation}\label{mtrcfcN2}
\begin{aligned}
    \horfc &= -X_{2} = (1-\lambda r^2)(a^2+r^2) - 2m (1+\lambda a^2)\, r\;,\\
    {\cal S} &= -X_{1} = (1+\lambda x^2)(a^2-r^2) - 2\alpha a (1+\lambda a^2)\, x\;,\\
    \Sigma &= \U_2 = - \U_1 = r^2 + x^2\;.
\end{aligned}
\end{equation}
This is the standard Kerr-NUT-(A)dS solution written in Carter's coordinates \cite{Carter:1968b, plebanski1975class}. It is labeled by parameter ${\lambda}$ fixed by the cosmological constant, mass ${m}$, NUT parameter ${\alpha}$, and rotation parameter ${a}$. The exact interpretration of parameters for nonvanishing ${\alpha}$ is, however, nontrivial, see \cite{GriffithsPodolsky:2006b, GriffithsPodolsky:book}.

We perform the limit ${N=2 \to \Nb=1,\,\Nt=1}$ which leads to the warped metric
\begin{equation}\label{mtrclimitwarpN2}
    \mtrc = \mtrt+\tilde{r}^2\mtrb|_{\lambda=1}\;.
\end{equation}
The regular sector is characterized by ${\Nt=1}$ and signature ${(-+)}$. Coordinates ${\tilde{r}}$, ${\tilde{t}}$ and mass parameter ${\tilde{m}}$ are trivially related to the original ones: ${\tilde{r}=r}$, ${\tilde{t}=t}$, and ${\tilde{m}=m}$.
The regular metric $\mtrt$ is given by
\ba
\mtrt&=&-\horfcd\grad\tilde{t}^2+\frac1\horfcd\grad\tilde{r}^2\;,\label{mtrctilN1}\\
\horfcd&=&1-\lambda r^2-\frac{2m}r\;.\label{horfctilN1}
\ea

The unspinned sector is characterized by ${\Nb=1}$, signature ${(++)}$, and the following notation dictionary: ${\xb_1\to\xb}$, ${\psb_0\to\psb}$, ${\ab_1\to\ab}$, ${\NUTb_1\to\bar{\alpha}}$. Coordinates and parameters are related to those before the limit as
${\xb\approx\frac1\zr x}$, ${\psb\approx-\zr\frac1{\lambda}\psi}$, ${\ab\approx\frac1\zr a}$, ${\bar{\alpha}=\alpha}$. The metric reads
\ba
\mtrb|_{\lambda=1}&=&\frac1\Xb\grad\xb^2+\Xb\grad\psb^2\;,\label{mtrcbarN1}\\
\Xb&=&\ab^2- \xb^2-2\bar\alpha \ab \xb\;.\label{mtrcfcbarN1}
\ea
By rescaling coordinates,
\begin{equation}\label{coorrescalingN1}
    \xi = \frac{\xb}{\ab}\;,\quad\phi=\ab\psb\;,
\end{equation}
it is possible to eliminate parameter ${\ab}$, obtaining the following metric (setting ${\bar\alpha=\alpha}$):
\ba
\mtrb|_{\lambda=1} &=& \frac1{X}\grad\xi^2+X\grad\phi^2\;,\label{mtrcbarN1xi}\\
 X &=& 1-\xi^2-2\alpha\xi\;.\label{DeltabarN1}
\ea
This metric is an Euclidean version of the 2-dimensional black hole solution of the $D\to2$ Einstein--dilaton gravity with a cosmological constant constructed in \cite{Mann:1989gh}, see also \cite{Frassino:2015oca} for the interpretation of parameters and its thermodynamics.

Coordinate ${\xi}$ runs between two roots of $X$, ${-\alpha}$ being the center of this interval. An introduction of the angular coordinate ${\tht\in(0,\pi)}$,
\begin{equation}\label{thtN1def}
  \xi = -\alpha+\sqrt{1+\alpha^2}\cos\tht\;,
\end{equation}
brings the metric to its spherical form
\begin{equation}\label{mtrcbarN1tht}
    \mtrb|_{\lambda=1} = \grad\tht^2+(1+{\alpha}^2)\sin^2\tht\,\grad\phi^2\;,
\end{equation}
with ${\alpha}$ affecting the angular deficit at poles of the sphere. We see that in this case, the NUT parameter ${\alpha}$ is rather trivial: it can be absorbed into conicity of the symmetry axis of the geometry.

Putting everything together, the obtained metric reads
\begin{equation}\label{mtrclimitD4}
    \mtrc = -\horfcd\grad\tilde{t}^2 + \frac1\horfcd\grad\tilde{r}^2
       +\tilde{r}^2 \bigl(\grad\tht^2+(1+\alpha^2)\sin^2\tht\,\grad\phi^2\bigr)\;,
\end{equation}
with the metric function $\horfcd$ given by \eqref{horfctilN1}. By switching-off the rotation we have recovered the Schwarzschild-(A)dS solution.

This examples illustrates, that although the NUT parameter $\alpha$ survived in this case, it belongs to the unspinned sector and hence completely decouples from temporal and radial coordinates. Therefore, after the limit it does not play a role of real NUT charge---we have not recovered the Taub--NUT solution \cite{taub1951empty, newman1963empty}. 
Since our limiting procedure completely decouples regular and unspinned sectors,
to maintain a nontrivial Lorentzian NUT parameter, we would need to have $\Nt\ge2$.
Even in the regular sector the surviving parameter $\alpha$ is rather trivial. For $\Nb=1$, it contributes only to the conicity of the geometry. This is not the case for all surviving NUT parameters in the unspinned sector in higher dimensions $\Nb>1$ as we are going to see in the next subsection.

\subsection{${D=6}$ dimensions}
\label{ssc:D6}

For the ${N=3}$ Lorentzian on-shell Kerr-NUT-(A)dS metric we employ the following notation dictionary: ${x_1,x_2\to x,y}$, ${\psi_0,\psi_1,\psi_2\to t,\ph,\psi}$, ${a_1,a_2\to a,b}$, and ${\NUT_1,\NUT_2\to\alpha,\beta}$. The metric \eqref{KerrNUTAdSphiWick} then reads
\begin{equation}\label{KerrNUTAdSN3}
\begin{split}
  \mtrc =
   & -\frac\horfc\Sigma\bigl(\grad t +(x^2+y^2)\grad\ph+x^2y^2\grad\psi\bigr)^2
     +\frac\Sigma\horfc\grad r^2\\
   & +\frac{U}{X}\grad x^2 + \frac{X}{U}\bigl(\grad t +(y^2-r^2)\grad\ph-y^2r^2\grad\psi\bigr)^2\\
   & +\frac{V}{Y}\grad y^2 + \frac{Y}{V}\bigl(\grad t +(x^2-r^2)\grad\ph-x^2r^2\grad\psi\bigr)^2\;,
\end{split}
\end{equation}
where
\begin{gather}\label{mtrcfcN3}
\begin{aligned}
    \horfc & = (1{-}\lambda r^2)(a^2{+}r^2)(b^2{+}r^2)
      -2m (1{+}\lambda a^2)(1{+}\lambda b^2)\,r\;,\\
    X &= -(1{+}\lambda x^2)(a^2{-}x^2)(b^2{-}x^2)
      +2\alpha a(1{+}\lambda a^2)(b^2{-}a^2)\,x\;,\\
    Y &= -(1{+}\lambda y^2)(a^2{-}y^2)(b^2{-}y^2)
      +2\beta b(1{+}\lambda b^2)(a^2{-}b^2)\,y\;,
\end{aligned}\notag\\[1ex]
\begin{aligned}
    \Sigma &= U_3 = (x^2+r^2)(y^2+r^2)\;,\\
    U &= U_1 = (x^2-y^2)(x^2+r^2)\;,\\
    V &= U_2 = (y^2-x^2)(y^2+r^2)\;.
\end{aligned}
\end{gather}

\subsubsection{Black hole with one rotation}
Let us first perform a single-spin-zero limit ${N=3\to \Nb=1,\,\Nt=2}$, switching-off one rotation parameter, ${a_1\to0}$. We obtain the warped metric, cf.~\eqref{mtrclimitsc},
\begin{equation}\label{mtrclimitwarpN3}
    \mtrc = \mtrt+\tilde{r}^2\frac{\tilde{y}^2}{\tilde{b}^2}\mtrb|_{\lambda=1}\;.
\end{equation}

In the regular sector we have a 4-dimensional metric ${\mtrt}$ with signature ${(-+++)}$ for which we employ the following notation dictionary: ${\xt_1\to\tilde{y}}$, ${\pst_1\to\tilde{\ph}}$, ${\at_1\to\tilde{b}}$, ${\NUTt_1\to\tilde{\beta}}$. As usual, in the regular sector coordinates and parameters after the limit are trivially related to those before the limit: ${\tilde{y}=y}$, ${\tilde{r}=r}$, ${\tilde{t}=t}$, ${\tilde{\ph}=\ph}$, ${\tilde{m}=m}$, ${\tilde{\beta}=\beta}$, and ${\tilde{b}=b}$. The metric resembles the Kerr-NUT-(A)dS solution \eqref{KerrNUTAdSN2} in four dimensions,
\begin{equation}\label{KerrNUTAdStilN2}
\begin{split}
  \mtrt =
    &-\frac{\horfc}{\tilde\Sigma}\bigl(\grad \tilde{t} + \tilde{y}^2\grad\tilde{\ph}\bigr)^2 + \frac{\tilde\Sigma}\horfc\,\grad \tilde{r}^2\\
    &+\frac{\tilde{\Sigma}}{\tilde\Delta}\,\grad \tilde{y}^2+\frac{\tilde\Delta}{\tilde\Sigma}\bigl(\grad \tilde{t} - \tilde{r}^2 \grad\tilde{\ph}\bigr)^2\;,
\end{split}
\end{equation}
however, the metric functions have the ${\Nb=1}$ off-shell form \eqref{Xtilded} instead of the usual \eqref{mtrcfcN2},
\begin{equation}\label{mtrcfctilN2}
\begin{aligned}
    \horfc &= -\Xt_{2} = (1-\lambda \tilde{r}^2)(\tilde{b}^2+\tilde{r}^2)
      - 2\tilde{m} (1+\lambda \tilde{b}^2) \frac1{\tilde{r}}\;,\\
    \tilde\Delta &= -\Xt_{1} = (1+\lambda \tilde{y}^2)(\tilde{b}^2-\tilde{y}^2)
      - 2\tilde{\beta} \tilde{b}^3 (1+\lambda \tilde{b}^2) \frac1{\tilde{y}}\;,\\
    \tilde\Sigma &= \Ut_2 = - \Ut_1 = \tilde{r}^2 + \tilde{y}^2\;.
\end{aligned}
\end{equation}

The unspinned sector leads to the ${\Nt=1}$ metric ${\mtrb|_{\lambda=1}}$ with signature ${(++)}$. With the notation dictionary ${\xb_1\to \bar{x}}$, ${\psb_0\to\bar\psi}$, ${\ab_1\to\bar{a}}$, and ${\NUTb_1\to\bar{\alpha}}$, it takes exactly the form \eqref{mtrcbarN1} studied in the previous section;  a discussion leading to \eqref{mtrcbarN1tht} applies again. New coordinates and parameters are related to those before the limit as follows: ${\bar{x}\approx\frac1\zr x}$, ${\bar{\psi}\approx-\zr\frac{b^2}{\lambda}\psi}$, ${\bar{a}\approx\frac1\zr a}$, and ${\bar{\alpha}=\alpha}$.

We can observe that the Killing coordinate ${\bar{\psi}}$ decoupled from the time coordinate ${\tilde{t}}$ (there are no off-diagonal terms ${g_{\tilde{t}\bar{\psi}}}$), i.e., there is no rotation in this direction. However, there remains a nontrivial rotation in the ${\tilde{\ph}}$ direction. The surviving NUT parameter $\tilde{\beta}$ belongs to the regular sector and as such it is coupled to the temporal coordinate. Consequently, its presence has familiar NUT-like consequences on the geometry, including the existence of closed time-like curves around the axis. The case with vanishing $\tilde{\beta}$ has been discussed in \cite{HawkingEtal:1999}.

\subsubsection{Twisted black hole}

Another alternative in six dimensions is to perform the multiple-spin-zero limit ${N=3\to \Nb=2,\,\Nt=1}$. The resulting metric reads  \eqref{mtrclimitNt=1},
\begin{equation}\label{mtrclimitN3Nt=1}
    \mtrc= -\horfcd\grad t^2 + \frac1\horfcd \grad r^2
       + r^2 \mtrb|_{\lambda=1}
\end{equation}
with the metric function\footnote{%
We have eliminated tildes in the regular sector in this case, since tilded quantities are trivially related to those before the limit: ${\tilde{r}=r}$, ${\tilde{t}=t}$, ${\tilde{m}=m}$.}
\begin{equation}\label{horfctilN3N1}
    \horfcd = 1-\lambda r^2 - \frac{2 m}{r^3}\;.
\end{equation}
We recognize the Schwarzschild--Tangherlini-like dependence on radial coordinate in six dimensions. However, the geometry reduces to the Schwarzschild--Tangherlini solution only for the vanishing NUT parameters in the unspinned sector.

Indeed, an unspinned metric ${\mtrb|_{\lambda=1}}$ is the ${\Nb=2}$ Euclidean instanton with signature ${(++++)}$. With the notation dictionary ${\xb_1,\xb_2\to \bar{x},\bar{y}}$, ${\psb_0,\psb_1\to\bar{\ph},\bar{\psi}}$, ${\NUTb_1,\NUTb_2\to\alpha,\beta}$, and ${\ab_1,\ab_2\to\bar{a},\bar{b}}$, it reads
\begin{equation}\label{EuclInstbarN2}
\begin{split}
 \mtrb|_{\lambda=1}
   &= \frac{\bar{y}^2-\bar{x}^2}{\bar{X}}\grad\bar{x}^2
     +\frac{\bar{X}}{\bar{y}^2-\bar{x}^2}\bigl(\grad\bar{\ph}+\bar{y}^2\grad\bar{\psi}\bigr)^2\\
   &+ \frac{\bar{y}^2-\bar{x}^2}{-\bar{Y}}\grad\bar{y}^2
     +\frac{-\bar{Y}}{\bar{y}^2-\bar{x}^2}\bigl(\grad\bar{\ph}+\bar{x}^2\grad\bar{\psi}\bigr)^2\;,
\end{split}
\end{equation}
where
\begin{equation}\label{mtrcfcEuclInstbarN2}
\begin{aligned}
  \bar{X} &= (\bar{a}^2-\bar{x}^2)(\bar{b}^2-\bar{x}^2)
      -2\bar{\alpha}\bar{a}(\bar{b}^2-\bar{a}^2)\,\bar{x}\;,\\
  \bar{Y} &= (\bar{a}^2-\bar{y}^2)(\bar{b}^2-\bar{y}^2)
      -2\bar{\beta}\bar{b}(\bar{a}^2-\bar{b}^2)\,\bar{y}\;.
\end{aligned}
\end{equation}
The ranges of ${\bar{x}}$ and ${\bar{y}}$ are given by intervals between the roots of ${\bar{X}}$ and  ${\bar{Y}}$, respectively, such that ${-\bar{a}<\bar{x}<\bar{a}<\bar{y}<\bar{b}}$, see the general discussion near \eqref{xrange}. The metric is not diagonal, it mixes Killing directions (we have a non-vanishing term ${g_{\bar{\ph}\bar{\psi}}}$), which we refer to as local twisting of the geometry.
For ${\bar{\alpha},\bar{\beta}>0}$, the curvature is singular only at ${r=0}$. For demonstration, the Kretschmann scalar is
\begin{equation}\label{KretschmannN3}
\begin{split}
    \mathcal{K} &= 100\lambda^2 + \frac{1600 m^2}{r^{10}}
    + \frac{80\bigl(\bar{a}^2-\bar{b}^2\bigr)^2}{r^4\bigl(\bar{y}^2-\bar{x}^2\bigr)^6}\\
    &\times\Bigl(
         (\bar{a}^2\bar{\alpha}^2+\bar{b}^2\bar{\beta}^2)(\bar{x}^2+\bar{y}^2)(\bar{x}^4+14\bar{x}^2\bar{y}^2+\bar{y}^4)\\
         &\qquad+4\bar{a}\bar{b}\bar{x}\bar{y}(3\bar{x}^2+\bar{y}^2)(\bar{x}^2+3\bar{y}^2)
         \Bigr)\;,
\end{split}
\end{equation}
${\bar{x}=\bar{y}}$ is out of the allowed coordinate range.
There can exist additional non-regularities on the axes of symmetry. They depend on a choice of identification of Killing coordinates, which also introduces global twisting of geometry in these directions.

The limiting metric \eqref{mtrclimitNt=1} thus represents a generalization of the static Schwarzschild--Tangherlini solution with angular part given by the twisted four-dimensional Euclidean instanton \eqref{EuclInstbarN2}.

\subsubsection{Deformed black hole}
After `decoupling' the 4-dimensional Euclidean instanton, by ``unspinning'' it from radial and temporal directions, we can also ``untwist'' its Killing angular direction. Performing so the second single-spin-zero limit $\Nb=2\to\bar{\Nb}=1,\,\tilde{\Nb}=1$, we arrive at the `combined metric' \eqref{limmtrc}. Using the following notations: $\xi_1,\xi_2\to\xi,\upsilon$, $\phi_1,\phi_2\to\phi,\psi$, $\NUT_1,\NUT_2\to\alpha,\beta$, the metric reads
\begin{equation}\label{mtrcmlimitN3}
\begin{split}
    \mtrc &= \mtrq_r + r^2\mtrq_\upsilon + r^2\upsilon^2 \mtrq_\xi \\
       &= -\horfcd\grad t^2 + \frac1\horfcd \grad r^2
             + r^2 \Bigl(\frac1{\Delta_\upsilon}\grad\upsilon^2+\Delta_\upsilon\grad\psi^2\Bigr)\\
       &\mspace{148mu}+ r^2\upsilon^2 \Bigl(\frac1{\Delta_\xi}\grad\xi^2+\Delta_\xi\grad\phi^2\Bigr)\;,
\end{split}
\end{equation}
where
\begin{equation}\label{mtrcfcmlimN3}
\begin{aligned}
    \horfcd &= 1-\lambda r^2 -\frac{2m}{r^3}\;,\\
    \Delta_\upsilon &= 1- \upsilon^2 - \frac{2\beta}\upsilon\;,\\
    \Delta_\xi &= 1- \xi^2 - 2\alpha\xi\;.
\end{aligned}
\end{equation}
Metric $\mtrq_r=-\horfcd\grad t^2+\horfcd^{-1}\grad r^2$ describes a geometry spanning the radial and temporal directions and has the familiar Schwarzschild--Tangherlini form. Metric ${\mtrq_\xi=\Delta_\xi^{-1}\grad\xi^2+\Delta_\xi\grad\phi^2}$ is spherical, \eqref{mtrcbarN1xi}. The middle part $\mtrq_\upsilon = \Delta_\upsilon^{-1}\grad\upsilon^2+\Delta_\upsilon\psi^2$ has the Euclidean signature for $\frac1{\sqrt{27}}>\beta\ge0$, when ${\Delta_\upsilon}$ has two roots which specify the allowed range of ${\upsilon}$. For ${\beta\neq0}$ this metric describes a non-spherical geometry.

The solution \eqref{mtrcmlimitN3} thus describes a static deformed Schwarzschild--Tangherlini black hole. Since the metric is diagonal, its Killing directions are not twisted. The curvature is singular for ${r=0}$ as can be seen, e.g., from the Kretschmann scalar
\begin{equation}\label{KretschmannN3multi}
    \mathcal{K} = 60\lambda^2 + \frac{960 m^2}{r^{10}} + \frac{48\beta^2}{r^4\upsilon^6}\;.
\end{equation}
Remember that for ${\beta>0}$, value ${\upsilon=0}$ is outside the allowed coordinate range.

\section{Discussion}
\label{sc:Discussion}

In the present paper we have analyzed basic properties of the higher-dimensional Kerr-NUT-(A)dS metrics. In particular, we discussed the signature and coordinate ranges, and identified the maximally symmetric space, the Euclidean instanton, and the black hole solution. We have then constructed a new class of black hole solutions of the Einstein equations with a cosmological constant by taking a special (singular) limit in the space of the Kerr-NUT-(A)dS metrics.

As shown by Geroch \cite{Geroch:1969}, limiting procedures of this kind are generally non-unique.  A procedure we adopted in the paper is singled out by the following properties. The Kerr-NUT-(A)dS metric in $2N$ dimensions possesses, besides the cosmological constant, $2N-1$ free physical parameters. The limiting metric has one parameter less. Geodesic equations in the Kerr-NUT-(A)dS geometry admit $2N$ independent integrals of motion in involution: $N$ of them are of the first order in momentum, while the other $N$ are quadratic in momentum. The limiting space preserves this property. Moreover, no new independent integrals of motion are in general present in the obtained space. The Kerr-NUT-(A)dS metric admits a non-degenerate CCKY tensor $\tens{h}$, responsible for the integrability properties of this space. In our limiting procedure this tensor becomes degenerate, $\tens{h}\approx \tens{\tilde h}$. As a result, the metric belongs to a wider class of metrics called the generalized Kerr-NUT-(A)dS spacetimes (see App.~A). Such metrics do not necessarily admit the full Killing tower of symmetries. However, our limiting procedure was constructed in such a way, that the adopted `enhancement' of the vanishing part of the CCKY tensor results in appearance of a new additional two-form $\tens{\bar h}$. This object `takes care' of the additional integrals of motion, which `disappeared' because of the degeneracy of the surviving CCKY tensor $\tens{\tilde h}$. In other words, the full Killing tower of symmetries still exists, though it cannot be generated from a single non-degenerate CCKY tensor.

The obtained black hole solutions generalize the Kerr-(A)dS spacetimes constructed in \cite{MyersPerry:1986, GibbonsEtal:2004,GibbonsEtal:2005} by including the Euclidean and Lorentzian NUT charges. The resulting geometry has a structure of warped space with the Kerr-like Lorentzian part warped to a Euclidean metric of deformed and/or twisted sphere, with the deformation and twist characterized by the Euclidean NUT parameters. The presence of Euclidean NUTs is less severe than that of their Lorentzian counterparts; typically Euclidean NUT charges smoothen the curvature singularity and introduce a global twisting of the geometry (the axes of symmetry may suffer from non-regularity due to identifications of Killing angles). As expected, for vanishing NUTs the metric reduces to the Kerr-(A)dS black hole spacetime, written in Jacobi-type coordinates \cite{Oota:2008uj}, with several rotations switched off.

In the present paper we have limited ourselves to even dimensions and the limit of vanishing rotation parameters. A few interesting possible developments immediately come to mind. The most obvious one is the extension of the present results to odd dimensions. Next is to consider another interesting limit where some of the rotation parameters are set equal but non-zero. One may expect that the resulting metric also belongs to the generalized Kerr-NUT-(A)dS class and possesses some enhanced symmetry. In particular, it is known that a very special subcase of such a limit leads to the `most general known' Einstein--K\"ahler manifolds \cite{Kubiznak:2009}. One might be even able to obtain the recently constructed super-entropic black holes \cite{Hennigar:2015cja}.

Although we studied the limit under which the NUT parameters survive and characterize a deformation of the resulting geometry, our limiting procedure is significantly different from the four-dimensional limit in which the Carter--Pleba\'{n}ski generic metric is turned to the Taub--NUT solution, see, e.g., \cite{GriffithsPodolsky:book}. In our limit, the metric functions $\mX{p}_\mu$ are approximately given by the even polynomials $\Ja(x_\mu^2)$ with roots $\pm a_\mu$, and these roots get only slightly modified by the linear NUT term. By sending $\ab_\mub\to0$ we send some of these roots to zero. As a result the spacetime assumes a warp structure with ``minimal coupling'' between the regular and unspinned sectors. On other hand, the Taub--NUT limit corresponds to the case when the even polynomial $\Ja(x_\mu^2)$ does not have maximal number of real roots and the real roots appear only due to the NUT term. The Taub--NUT solution is then obtained when two of these roots coincide. A detailed discussion of such a limit in higher dimensions awaits to be properly studied.

Another future direction would be to generalize the above limit to solutions with non-trivial matter fields. In particular, one may consider a test electromagnetic field \cite{FrolovKrtous:2011} and its impact on integrability of charged geodesic motion.

Surprisingly, not much is still known about the higher-dimensional Kerr-NUT-(A)dS geometry. Being a higher-dimensional generalization of the Carter--Pleba\'{n}ski class it likely contains a large family of solutions of Einstein's equations of various signatures and physical interpretations. In the present paper we have slightly uncovered the basic structure of this family. In particular, by studying the signature and ranges of coordinates we identified black hole solutions of the Kerr-(A)dS type. We also were able to find one special subclass contained in this family that arises from switching off one of the `rotation parameters'. However, many open tasks remain. Probably the most difficult is to analyze the meaning of free parameters that appear in the Kerr-NUT-(A)dS solution. Analogous task has been accomplished for the Carter--Pleba\'{n}ski class only recently \cite{GriffithsPodolsky:2006b,GriffithsPodolsky:book}. The process of identification of these parameters is intrinsically related to a procedure by which various special limits can be taken and various subclasses identified. The metrics of the generalized Kerr-NUT-(A)dS type are likely to emerge in this process, smearing the gap between the two families. We believe that, similar to its four-dimensional Carter--Pleba\'{n}ski counterpart, the (generalized) Kerr-NUT-(A)dS family `hides' many new interesting solutions yet to be discovered.

\section*{Acknowledgments}

V.F.\ thanks the Natural Sciences and Engineering Research Council of Canada and the Killam Trust for financial support.
P.K.\ was supported by Grant GA\v{C}R~P203/12/0118.
D.K. is supported in part by Perimeter Institute for Theoretical Physics and by the Natural Sciences and Engineering Research Council of Canada. Research at Perimeter Institute is supported by the Government of Canada through Industry Canada and by the Province of Ontario through the Ministry of Research and Innovation.
I.K.\ was supported by the Charles University Grant No. SVV-260211.
V.F.\ and P.K.\ appreciate the hospitality of the Perimeter Institute, Waterloo, where this work started.

\appendix

\section{Generalized Kerr-NUT-(A)dS spacetimes}\label{apx:degPCCKYfam}

The generalized Kerr-NUT-(A)dS metric is the most general metric that admits a (not necessarily non-degenerate) rank-2 CCKY tensor \cite{HouriEtal:2008b,HouriEtal:2009, Oota:2008uj}. In a general dimension $D$, the metric possesses a bundle structure, with the fiber being the $2n$-dimensional Kerr-NUT-(A)dS metric and the base taking a form of the product space $B=M^1\times M^2\times\dots M^I\times M^0$. Here the manifolds $M^i$ are $2m_i$-dimensional K\"ahler manifolds with metrics $\tens{g}^i$ and K\"ahler 2-forms $\tens{\omega}^i=\tens{dB}^i$, and $M^0$ is an `arbitrary' manifold of dimension $m_0$ and metric $\tens{g}^0$,
\be
D=2n+2|m|+m_0\,,\quad |m|=\sum_{i=1}^I m_i\,.
\ee
The generalized Kerr-NUT-(A)dS metric takes the following form:
\ba
\tens{g}&=&\sum_{\mu=1}^n \frac{\tens{d}x_\mu^2}{P_\mu(x)}+\sum_{\mu=1}^n P_\mu(x)\Bigl(\sum_{k=0}^{n-1}A_\mu^{(k)}\tens{\theta}_k\Bigr)^2\nonumber\\
&&+\sum_{i=1}^I\prod_{\mu=1}^n(x_\mu^2-\xi_i^2)\tens{g}^i+A^{(n)}\tens{g}^0\,,
\ea
while the CCKY 2-form reads
\be
\tens{h}=\sum_{\mu=1}^n x_\mu \tens{d} x_\mu\wedge\Bigl(\sum_{k=1}^{n-1}A_\mu^{(k)}\tens{\theta}_k\Bigr)
+\sum_{i=1}^I \xi_i \prod_{\mu=1}^n (x_\mu^2-\xi_i^2)\tens{\omega}^i\,,
\ee
where
\ba
\tens{\theta}_k&=&\tens{d}\psi_k-2\sum_{i=1}^I(-1)^{n-k}\xi_i^{2(n-k)-1}\tens{B}^i\,,\nonumber\\
P_\mu&=&X_\mu(x_\mu)\Bigl[x_\mu^{{m}_0}\prod_{i=1}^I(x_\mu^2-\xi_i^2)^{m_i}(-1)^nU_\mu\Bigr]^{-1}\,.
\ea
Coordinates $x_\mu$ are the non-constant functionally independent eigenvalues of $\tens{h}$, whereas parameters $\xi_i$ stand for the non-zero
constant eigenvalues of $\tens{h}$, each having multiplicity $m_i$ that determines the dimension of K\"ahler manifolds $M^i$. The dimension
$m_0$ of the manifold $M^0$ equals the multiplicity of the zero value eigenvalue of $\tens{h}$.
For $m_0=1$, the metric $\tens{g}^0$ can take a special form
\be\label{special}
A^{(n)}\tens{g}^0=\frac{c}{A^{(n)}}\Bigl(\sum_{k=0}^nA^{(k)}\tens{\theta}_k\Bigr)^2\,.
\ee
Assuming that the base metrics $\tens{g}^0$ and $\tens{g}^i$ are Einstein spaces with cosmological constants $\lambda^0$ and $\lambda^i$, the generalized Kerr-NUT-(A)dS metric solves the vacuum Einstein equations  with the cosmological constant, $\Ric_{ab}=\lambda \tens{g}_{ab}$,
provided the metric functions $X_\mu$ take the following form:
\be
X_\mu=x_\mu\Bigl(d_\mu+\int\chi(x_\mu)x_\mu^{m^0-2}\prod_{i=1}^I(x_\mu^2-\xi_i^2)^{m_i}dx_\mu\Bigr)\,,
\ee
where
\be
\chi(x)=\sum_{k=-\epsilon}^n\alpha_i x^{2i}\,,\quad \alpha_n=-\lambda\,.
\ee
For the general type ($\epsilon=0$) of $\tens{g}^0$, we have $\alpha_0=(-1)^{n-1}\lambda^0$, whereas for the special type
of $\tens{g}^0$ described by \eqref{special}  we have ($\epsilon=1$)
\be
\alpha_0=(-1)^{n-1}2c\sum_{i=1}^I\frac{m_i}{\xi_i^2}\,,\quad \alpha_{-1}=(-1)^{n-1}2c\,.
\ee
The cosmological constants $\lambda^i$ are given by $\lambda^i=(-1)^{n-1}\chi(\xi_i)$.

A subfamily of solutions with vanishing NUT charges, describing the Kerr-(A)dS black holes with
partially equal and some vanishing angular momenta have been identified in \cite{Oota:2008uj}. The corresponding scaling limit
is in a way analogous to the one performed in the main text of this paper, with the exception that in \cite{Oota:2008uj} the NUT parameters were
switched off.

Let us stress that the generalized Kerr-NUT-(A)dS metrics do not necessarily admit the Killing tower of symmetries. The presence of a degenerate CCKY tensor is not enough to generate this full tower and much smaller subset of symmetries exists in these spacetimes. In particular, metrics $\tens{g}^i$ are in general `arbitrary' K\"ahler metrics without any additional symmetries.

The new metrics obtained in this paper, describing the twisted and deformed black holes, belong to the class of generalized Kerr-NUT-(A)dS spacetimes. In particular, they correspond to the even-dimensional case for which all the K\"ahler metrics $\tens{g}^i$ identically vanish and metric $\tens{g}^0$ becomes again the Kerr-NUT-(A)dS spacetime. As discussed in the main text, these solutions are distinguished by the presence of a full Killing tower of symmetries. Moreover, the presence of NUT charges makes them a non-trivial generalization of Kerr-(A)dS black holes discovered in
\cite{MyersPerry:1986, GibbonsEtal:2004,GibbonsEtal:2005}.

\section{Technical results}
\label{apx:tech}

\subsection{Properties of metric functions}
\label{apx:JAU}

In Sec.~\ref{sc:KerrNutAdS} we have defined auxiliary functions $\J(a^2)$ and $\A{k}$ as follows\footnote{%
Let us remind that, if not indicated otherwise, the sums (and products) run over ``standard'' ranges of indices:
\[
\sum_\mu \equiv \sum_{\mu=1}^N\;,\qquad
\sum_k \equiv \sum_{k=0}^{N-1}\;.
\]}
\begin{equation}\label{metricpolysx}
    \J(a^2)=\prod_{\nu}(x_\nu^2-a^2) =\sum_{k=0,\dots,N} \A{k} (-a^2)^{N{-}k}\;,
\end{equation}
and their complementary functions
\begin{equation}\label{metricpolysa}
    \Ja(x^2)=\prod_{\nu}(a_\nu^2-x^2) =\sum_{k=0,\dots,N} \Aa{k} (-x^2)^{N{-}k}\;.
\end{equation}
The above definitions imply that
\begin{align}
    \A{k} &=\sum_{\substack{\mu_1,\dots,\mu_k\\\mu_1<\dots<\mu_k}} x_{\mu_1}^2\dots x_{\mu_k}^2\;,\label{Ax}\\
    \Aa{k} &=\sum_{\substack{\mu_1,\dots,\mu_k\\\mu_1<\dots<\mu_k}} a_{\mu_1}^2\dots a_{\mu_k}^2\;.\label{Aa}
\end{align}
Similarly, we define functions $\J_\mu(a^2)$, $\A{j}_\mu$, $\Ja_\mu(x^2)$, and $\Aa{j}_\mu$, which skip the $\mu$th variables $x_\mu$ and $a_\mu$ as follows
\begin{align}
    \J_\mu(a^2)&=\prod_{\substack{\nu\\\nu\neq\mu}}(x_\nu^2-a^2)
       =\sum_{k} \A{k}_\mu (-a^2)^{N{-}1{-}k}\;,\label{metricpolysmux}\\
    \Ja_\mu(x^2)&=\prod_{\substack{\nu\\\nu\neq\mu}}(a_\nu^2-x^2)
       =\sum_{k} \Aa{k}_\mu (-x^2)^{N{-}1{-}k}\;,\label{metricpolysmua}
\end{align}
with
\begin{align}
    \A{k}_\mu &=\sum_{\substack{\nu_1,\dots,\nu_k\\\nu_1<\dots<\nu_k\\\nu_i\neq\mu}}
        x_{\nu_1}^2\dots x_{\nu_k}^2\;,\label{Amux}\\
    \Aa{k}_\mu &=\sum_{\substack{\nu_1,\dots,\nu_k\\\nu_1<\dots<\nu_k\\\nu_i\neq\mu}}
        a_{\nu_1}^2\dots a_{\nu_k}^2\;.\label{Amua}
\end{align}
These functions satisfy
\begin{equation}\label{J0}
\begin{gathered}
    \J(x_\mu^2)=0\;,\qquad  \Ja(a_\mu^2)=0\;,\\
    \J_\mu(x_\nu^2)=0\;,\qquad  \Ja_\mu(a_\nu^2)=0\;,\qquad\text{for $\nu\neq\mu$}\;.
\end{gathered}
\end{equation}
Finally, we define
\begin{align}
    \U_\mu = \J_\mu(x_\mu^2)= \prod_{\substack{\nu\\\nu\neq\mu}}(x_\nu^2-x_\mu^2)\;,\label{Uxdef}\\
    \Ua_\mu = \Ja_\mu(a_\mu^2)= \prod_{\substack{\nu\\\nu\neq\mu}}(a_\nu^2-a_\mu^2)\;.\label{Uadef}
\end{align}
The polynomials $\A{k}$ and $\A{k}_\mu$ satisfy the following identities:

\begin{equation}\label{Aid0}
    \A{k}=\A{k}_\mu+x_\mu^2\A{k{-}1}_\mu\;,
\end{equation}
\begin{gather}
\label{Aid1}
    \sum_k\A{k}_\mu \frac{(-x_\nu^2)^{N{-}1{-}k}}{\U_\nu}=\delta^\nu_\mu\;,\\
\label{Aid1i}
    \sum_\mu\A{k}_\mu \frac{(-x_\mu^2)^{N{-}1{-}l}}{\U_\mu}=\delta^k_l\;,\\
\label{id2}
    \sum_\mu\A{k}_\mu \frac{(-x_\mu^2)^{N}}{\U_\mu}=-\A{k{+}1}\;,\\
\label{id3}
    \sum_\mu\A{k}_\mu = (N-k)\A{k}\;,\\
\label{id4}
    \sum_\mu(N-k)\A{k} \frac{(-x_\nu^2)^{N{-}1{-}k}}{\U_\nu} = 1\;,\\
\label{Aid5}
    \sum_{k=0,\dots,N} \A{k} (-x_\nu^2)^{N{-}k} = 0\;,\\
\label{Aid6}
    \sum_{l=0,\dots,k} \A{l} (-x_\nu^2)^{k{-}l} = \A{k}_\mu\;.
\end{gather}
Analogous identities hold also for the complementary polynomials $\Aa{k}$ and $\Aa{k}_\mu$.

For the functions $\J(a^2)$ and $\Ja(x^2)$ we can write
\begin{equation}\label{Jid1}
\begin{gathered}
    \prod_\mu \J(a_\nu^2) = (-1)^N \prod_\nu \Ja(x_\mu^2)\;,\\
    \prod_{\substack{\mu\\\mu\neq\kappa}} \J_\kappa(a_\nu^2) =
       (-1)^{N{-}1} \prod_{\substack{\nu\\\nu\neq\kappa}} \Ja_\kappa(x_\mu^2)\;.
\end{gathered}
\end{equation}
These functions satisfy important orthogonality relations
\begin{equation}\label{Jortrel1}
    \sum_\alpha \frac{\J_\nu(a_\alpha^2)}{\Ua_\alpha}\frac{\Ja_\alpha(x_\mu^2)}{\U_\mu} = \delta^\mu_\nu\;,
\end{equation}
\begin{equation}\label{Jortrel2}
    \sum_\alpha \frac{\J_\mu(a_\alpha^2)\J_\nu(a_\alpha^2)}{\J(a_\alpha^2)\Ua_\alpha} =
      - \frac{\U_\mu}{\Ja(x_\mu^2)} \delta_{\mu\nu}\;,
\end{equation}
\begin{equation}\label{Jortrel3}
    \sum_\mu \J_\mu(a_\alpha^2)\J_\mu(a_\beta^2)\frac{\Ja(x_\mu^2)}{\U_\mu} =
      - \J(a_\alpha^2) \Ua_\alpha \delta^{\alpha\beta}\;.
\end{equation}

\subsection{The curvature tensor}\label{KretschSec}

The Riemann tensor of the off-shell Kerr-NUT-(A)dS metric has been calculated in \cite{HamamotoEtal:2007}. The obtained limiting metric \eqref{mtrclimit} has a warped-product structure with both components belonging to the Kerr-NUT-(A)dS class. The curvature of this metric can thus be obtained by using general formulae relating the curvature of the warped metric to curvatures of its components and derivatives of the warped factor. Following \cite{DobarroEtal:2004,KrtousKubiznakKolar:2015}, the Riemann and Ricci tensors read
\ba
\Riem_{ab}{}^c{}_d
    &=&\Riemt_{ab}{}^c{}_d+\frac{2}{\warp}\Hess_{d[a}\dltb^c_{b]}-2\warp\,\mtrt^{ce}\Hess_{e[a}\mtrb_{b]d}\qquad\nonumber\\
    &&+\Riemb_{ab}{}^c{}_d-2\warp^2\tilde{\lambda}^2\dltb^c_{[a}\mtrb_{b]d}\;,\label{eq:warpRiem}\\
\Ric_{ab}
    &=&\Rict_{ab}-\frac{\Nb}{\warp}\Hess_{ab}\nonumber\\
	 &&+\Ricb_{ab}-\warp^2\biggl(\frac{\hess}{\warp}+(\Nb{-}1)\tilde{\lambda}^2\biggr)\mtrb_{ab}\;.\qquad\label{eq:warpRic}
\ea
Scalar curvature $\scc$ and the Kretschmann scalar $\mathcal{K}$ are
\begin{gather}	
  \scc=\scct+\frac{\sccb}{\warp^2}-\frac{2\Nb}{\warp}\hess-\Nb(\Nb-1)\tilde{\lambda}^2\;,\label{eq:warpscc}\\
  \mathcal{K}
    =\tilde{\mathcal{K}}+\frac{4\Nb}{\warp^2}\Hess^{ab}\Hess_{ab}
	+\frac{1}{\warp^4}\bar{\mathcal{K}}
    -4\frac{\tilde{\lambda}^2}{\warp^2}\sccb
    +2(\Nb{-}1)\Nb\tilde{\lambda}^4\;.\label{eq:warpKret}
\end{gather}
Here, the Hessian tensor $\Hess$ and scalar $\hess$ are defined as
\begin{equation}\label{eq:hessian}
	\Hess_{ab} =\covdt_{\!a}\covdt_{\!b}\warp\;,\quad
	\hess = \mtrt^{ab}\,\Hess_{ab}\;,
\end{equation}
and
\begin{equation}\label{eq:lambda}
	\tilde{\lambda}^2 = \warp^{-2}\,\mtrt^{ab}\, \grad_a\warp\, \grad_b\warp\;.
\end{equation}
In particular, for the warp factor \eqref{warpdef} the Hessian tensor and scalars $\hess$, $\tilde{\lambda}^2$ can be expressed as
\begin{equation}\label{eq:auxquancoor}
\begin{gathered}
	\Hess = \sum_{\mut} \frac{\warp}{2\xt_{\mut}} \Bigl(\frac{\Xt_\mut}{\Ut_\mut}\Bigr)_{\!\!,\mut}
       \bigl(\enft{\mut}\enft{\mut}+\ehft{\mut}\ehft{\mut}\bigr)\;,\\
	\hess = \sum_{\mut}\frac{\warp\Xt_{\mut}'}{\xt_{\mut}\Ut_{\mut}}\;,\quad
	\tilde{\lambda}^2 =\sum_{\mut}\frac{\Xt_{\mut}}{\xt_{\mut}^2\Ut_{\mut}}\;.
\end{gathered}
\end{equation}

\subsection{Multiple-spin-zero limit: survival of the Killing tower}
\label{apx:KillTowLim}

Let us now provide more details about the multiple-spin-zero limit of the Killing tower discussed briefly in Sec.~\ref{LimitKTower}.
We mentioned there that although the limit of the CCKY 2-form $\tens{h}\approx \tens{\tilde h}$ becomes degenerate, the full Killing tower of symmetries nevertheless survives. In some sense this is caused by the emergence of a new \mbox{2-form} $\tens{\bar h}$ that is non-degenerate on manifold $\bar{\mathcal{M}}$. Let us first briefly explain how this additional 2-form emerges.

We start from the CCKY 2-form $\tens{h}$. In the limit $\zr\to 0$ its exterior power $\PCCKY^{\wedge \Nt}$ becomes a rank $2\Nt$ antisymmetric tensor that vanishes in the directions tangent to $\bar{\mathcal M}$. It can be also understood as a tensor on the submanifold $\tilde{\mathcal{M}}$. On this submanifold we can define the Levi-Civita tensor $\LCTt$ and, obviously, $\PCCKY^{\wedge \Nt}$ must be proportional to it with some suitable prefactor,
\begin{equation}
   \frac1{\Nt!} \PCCKY^{\wedge \Nt} \approx \frac1{\Nt!}\PCCKYt^{\wedge \Nt} = \sqrt{\At{\Nt}}\, \LCTt\;.
\end{equation}
One also has
\begin{equation}
  \frac1{(\Nt{+}1)!} \PCCKY^{\wedge (\Nt{+}1)} \approx \sqrt{\At{\Nt}}\, \LCTt \wedge  \PCCKY \approx \mathcal{O}(\zr)\;.
\end{equation}
As we will show, if this object is rescaled by $\zr^{-1}$, its limit has a structure of the wedge-product of tensors from $\tilde{\mathcal{M}}$ and $\bar{\mathcal{M}}$
\begin{equation}
    \PCCKY^{\wedge (\Nt{+}1)} \propto \LCTt\wedge \PCCKYb\;,
\end{equation}
cf.\ relations \eqref{CCKYlimit} for $\kb=1$. This determines a new \mbox{2-form} $\PCCKYb$ on $\mathcal{M}$ which is non-degenerate on the unspinned manifold $\bar{\mathcal{M}}$. In some sense this \mbox{2-form} is responsible for the survival of the Killing tower.

More concretely, as a result of the limiting procedure of the original CCKY tensor $\tens{h}$ one obtains a CCKY tensor ${\PCCKYt}$ and a new rank two antisymmetric tensor $\PCCKYb$. Although $\PCCKYt$ is degenerate on the full manifold, it is non-degenerate on $\tilde{\mathcal{M}}$ where it plays a role similar to $\tens{h}$ on $\mathcal{M}$. The other tensor, $\PCCKYb$, is not a CCKY 2-form of the limiting metric (since we factorized out some nontrivial prefactors), however we will show that it can effectively play a role of a non-degenerate CCKY tensor on $\bar{\mathcal{M}}$. Indeed, as discusses in the main text, the limiting geometry has a structure of the warped product \eqref{mtrclimit} and $\PCCKYt$ is the CCKY form of metric $\mtrt$ and $\PCCKYb$ is the CCKY form of metric $\mtrb$; both metrics belonging to the off-shell Kerr-NUT-(A)dS class. This guarantees the survival of the Killing tower.

With the above intuitive explanation in mind let us proceed to the actual limit of the Killing objects. The CCKY forms ${\CCKY{k}}$ defined in \eqref{CCKYdef} can be written as
\begin{equation}\label{CCKYomega}
    \CCKY{k} = {\frac1{\sqrt{\Aa{k}}}}
    \sum_{\mu_1<\dots<\mu_k} x_{\mu_1}\dots x_{\mu_k} \;\plnf{\mu_1}\wedge \dots\wedge\plnf{\mu_k}\;.
\end{equation}
For ${k=\kt\leq\Nt}$, the leading order contains just unscaled terms ${x_{\mu_i}}$ with ${\mu_i=\Nb+\mut_i>\Nb}$,
\begin{equation}\label{CCKYomegat}
    \CCKY{\kt} =  {\frac1{\sqrt{\Aa{k}}}}
    \sum_{\mut_1<\dots<\mut_\kt} \xt_{\mut_1}\dots \xt_{\mut_k} \;\plnft{\mut_1}\wedge \dots\wedge\plnft{\mut_k}\;,
\end{equation}
which together with \eqref{Alimit} proves the first relation \eqref{CCKYlimit}.

For ${k=\Nt+\kb>\Nt}$ the leading term in \eqref{CCKYomega} must contain ${\kb}$ rescaled terms, since there are just ${\Nt}$ unscaled coordinates ${x_{\Nb+1},\,\dots,\, x_{N}}$. The unscaled terms can be factorized in front of the sum and with the help of \eqref{Alimit}, \eqref{plnflimit}, and, in the second equality, with the help of \eqref{warpdef}, \eqref{mtrcLCT}, and again \eqref{CCKYomega}, we obtain
\begin{equation}\label{CCKYomegab}
\begin{split}
    \CCKY{\Nt{+}\kb}
    &\approx \zr^{-\kb} {\textstyle\bigl(\frac1{\Aat{\Nt}\Aab{\kb}}\bigr)^{\!\!\frac12}}
       \zr^\kb \warp^{2\kb}\; \xt_1\dots\xt_\Nt\;\plnft{1}\wedge\dots\wedge\plnft{\Nt}\\
    &\quad\qquad\qquad \wedge  \sum_{\mub_1<\dots<\mub_\kb} \xb_{\mub_1}\dots \xb_{\mub_\kb} \;\plnfb{\mub_1}\wedge \dots\wedge\plnfb{\mub_\kb}\\
    &= \warp^{2\kb{+}1}\,\LCTt\wedge \CCKYb{\kb}\;,
\end{split}\raisetag{6ex}
\end{equation}
which concludes the proof of relations \eqref{CCKYlimit}.

The limit of KY forms ${\KY{k}}$ can be obtained employing definition \eqref{KYdef}. The Hodge dual is given by the contraction with the Levi-Civita tensor ${\LCT}$ which splits into unscaled and scaled parts \eqref{LCTlimit}. For ${k=\kt\leq\Nt}$ we use the first relation of \eqref{CCKYlimit} and obtain
\begin{equation}\label{KYlimitproof1}
   \KY{\kt}
     =  \CCKY{\kt} \bullet \LCT\\
     \approx    \warp^{2\Nb} \CCKYt{\kt}\bullet (\LCTt\wedge\LCTb)\;.
\end{equation}
Since ${\CCKYt{\kt}}$ is nontrivial only in the unscaled directions, we can write
\begin{equation}\label{KYlimitproof2}
    \KY{\kt} \approx    \warp^{2\Nb} \bigl(\CCKYt{\kt}\bullett \LCTt\bigr)\wedge\LCTb
    =          \warp^{2\Nb} \,\KYt{\kt}\wedge\LCTb\;,
\end{equation}
which is the first relation \eqref{KYlimit}.

For ${k=\Nt+\kb>\Nt}$ we employ the second relation \eqref{CCKYlimit}, to get
\begin{equation}\label{KYlimitproof3}
\begin{split}
   &\KY{\Nt{+}\kb}
       = \CCKY{\Nt{+}\kb} \bullet \LCT\\
   &\quad\approx    \Bigl(\warp^{2\kb{+}1} \,\LCTt\wedge \CCKYb{\kb}\Bigr)
      \bullet (\warp^{2\Nb}\,\LCTt\wedge\LCTb)\\
   &\quad\approx    \warp^{2\kb{+}1} \warp^{2\Nb} \warp^{-4\kb}\,
      \bigl( \LCTt \bullett \LCTt\bigr) \Bigl( \CCKYb{\kb} \bulletb \LCTb\Bigr)\;.
\end{split}
\end{equation}\\[-1ex]
The additional warp-factor ${\warp^{-4\kb}}$ has appeared due to a conversion of ${\bullet}$ into ${\bulletb}$, since these operations involve the inverse metric ${\mtrc^{-1}}$, ${\mtrb^{-1}}$, respectively, used for raising ${\kb}$ indices before the contraction. Since ${\LCTt\bullett\LCTt=1}$, and recalling \eqref{KYdef}, we proved \eqref{KYlimit},
\begin{equation}\label{KYlimitproof4}
   \KY{\Nt{+}\kb}
   \approx   \warp^{2(\Nt{-}\kb){+}1}\, \KYb{\kb}\;.
\end{equation}

To find the limit of the Killing tensors \eqref{KTcoor},  it is useful to write them with indices in the lower position,
\begin{equation}\label{KTlower}
    \Kt{j} 
    =\sum_\mu \frac{\A{j}_\mu}{\Aa{j}} \Biggl[\,\frac{\U_\mu}{X_\mu}\,\grad x_\mu^2
       +\frac{X_\mu}{\U_\mu}\biggl(\sum_k \A{k}_\mu\grad{\psi_k}\biggr)^{\!\!2}\,\Biggr]\;.
\end{equation}
Employing relations \eqref{Alimit} and \eqref{mtrctermslimit}, the sum splits into sums over scaled and unscaled coordinates. For ${j=\jt\leq\Nt}$ both parts contribute to the leading order:

\begin{widetext}
\begin{equation}\label{KTlimitproof1}
\begin{split}
    \Kt{\jt}
    &\approx \frac1{\Aat{\jt}} \sum_\mut \At{\jt}_\mut
       \Biggl[\,\frac{\Ut_\mut}{\Xt_\mut}\,\grad \xt_\mut^2
       +\frac{\Xt_\mut}{\Ut_\mut}\biggl(\sum_\kt \At{\kt}_\mut\grad{\pst_\kt}\biggr)^{\!\!2}\,\Biggr]\\
    &+ \frac{\At{\jt}}{\Aat{\jt}} \sum_\mub
       \Biggl[\frac1{\zr^2}\frac{\At{\Nt}}{\Aat{\Nt}}\frac{\Ub_\mub}{\Xb_\mub}\,\zr^2\grad \xb_\mub^2
       +\zr^2\frac{\Aat{\Nt}}{\At{\Nt}}\frac{\Xb_\mub}{\Ub_\mub}\frac1{\zr^2}
       \biggl(\frac{\At{\Nt}}{\Aat{\Nt}}\sum_\kb \Ab{\kb}_\mub\grad{\psb_\kb}\biggr)^{\!\!2}\,\Biggr]\\
    &= \Ktt{\jt} + {\frac{\At{\jt}}{\Aat{\jt}}}\, \warp^{2}\, \mtrb\;.
\end{split}
\end{equation}
For ${j=\Nt+\jb>\Nt}$ the sum over unscaled coordinates disappears in the leading order:
\begin{equation}\label{KTlimitproof2}
\begin{split}
    \Kt{\Nt{+}\jb}
    &\approx \zr^{-2\jb}\frac1{\Aat{\Nt}\Aab{\jb}} \zr^{2(\jb{+}1)} \Ab{\jb{+}1}
       \sum_\mut \At{\Nt{-}1}_\mut
       \Biggl[\,\frac{\Ut_\mut}{\Xt_\mut}\,\grad \xt_\mut^2
       +\frac{\Xt_\mut}{\Ut_\mut}\biggl(\sum_\kt \At{\kt}_\mut\grad{\pst_\kt}\biggr)^{\!\!2}\,\Biggr]\\
    & +\zr^{-2\jb}\frac1{\Aat{\Nt}\Aab{\jb}} \zr^{2\jb} \At{\Nt}
       \sum_\mub \Ab{\jb}_\mub
       \Biggl[\frac1{\zr^2}\frac{\At{\Nt}}{\Aat{\Nt}}\frac{\Ub_\mub}{\Xb_\mub}\,\zr^2\grad \xb_\mub^2
       +\zr^2\frac{\Aat{\Nt}}{\At{\Nt}}\frac{\Xb_\mub}{\Ub_\mub}\frac1{\zr^2}
       \biggl(\frac{\At{\Nt}}{\Aat{\Nt}}\sum_\kb \Ab{\kb}_\mub\grad{\psb_\kb}\biggr)^{\!\!2}\,\Biggr]\\
    &\approx \warp^{4}\, \Kt{\jb}\;.
\end{split}
\end{equation}
\end{widetext}
Here, the functions ${\Xt_\mut}$ and ${\Xb_\mub}$ are given by relations \eqref{Xtilded} and \eqref{Xbarred}. Raising indices with the limiting metric \eqref{mtrclimit} we obtain relations \eqref{KTlimit}.

\subsection{A successive application of the limit}
\label{apx:multilimit}
In this appendix, we provide some details about the limiting procedure performed in Sec.~\ref{sc:MultiLimit}. Namely, we will repeatedly apply the multiple-spin-zero limit to the Kerr-NUT-(A)dS metric and its Killing tower objects.

Let us first list some useful relations for the intermediate metrics introduced in \eqref{omtrLCT} and \eqref{dmtrLCT}. Metric ${\mtro}$ and its Levi-Civita tensor can be split into a warped product as follows
\begin{equation}\label{limmtrcwarppr}
\begin{aligned}
  \mtro &= {}^\alpha\mtrd + {}^\alpha w^2\; {}^{N{-}\alpha}\mtro\;,\\
  \LCTo &= {}^\alpha\LCTd \wedge {}^\alpha w^{2(N{-}\alpha)}\; {}^{N{-}\alpha}\LCTo\;,
\end{aligned}
\end{equation}
where, for arbitrary $\alpha$, the warp factor reads
\begin{equation}\label{warplim}
    {}^\alpha w = \sx_N\dots\sx_{N{-}\alpha{-}1}\;.
\end{equation}
Intermediate metrics \eqref{omtrLCT} and \eqref{dmtrLCT} satisfy recurrence relations
\begin{equation}\label{recrel}
\begin{gathered}
  {}^{\alpha{+}1}\mtro = \mtrq_{\alpha{+}1} + \sx_{\alpha{+}1}^2\; {}^\alpha\mtro\;,\quad
  {}^{\alpha{+}1}\LCTo = \LCTq_{\alpha{+}1} \wedge \sx_{\alpha{+}1}^{2\alpha}\; {}^\alpha\LCTo\;,\\
  {}^{\alpha{+}1}\mtrd = {}^\alpha\mtrd + {}^\alpha w^2\; \mtrq_{N{-}\alpha}\;,\quad
  {}^{\alpha{+}1}\LCTd = {}^\alpha\LCTd \wedge {}^\alpha w^2\; \LCTq_{N{-}\alpha}\;.
\end{gathered}
\end{equation}

We want to take repeatedly the limit ${\Nb=N-1}$, ${\Nt=1}$. To obtain the limit of the metric, it is enough to look at one ``inductive'' step.
We claim that after ${N-\alpha}$ limiting steps, the metric is
\begin{equation}\label{mtrclimstep1}
   \mtrc \approx {}^{N-\alpha}\mtrd + {}^\alpha w^2\; {}^\alpha\mtrc\;.
\end{equation}
Let us assume it for a particular ${\alpha}$. Performing the ${\Nt=1}$ limit \eqref{mtrclimit} of the metric ${{}^\alpha\mtrc}$ we get
\begin{equation}\label{mtrclimstep2}
   \mtrc \approx {}^{N-\alpha}\mtrd + {}^\alpha w^2\;
     \bigl({}^{\alpha}\mtrt + \frac{\xt_{1}^2}{\at_1^2}\; {}^{\alpha}\mtrb\bigr)\;.
\end{equation}
Introducing the rescaled coordinate ${\sx_\alpha=\frac{\xt_1}{\at_1}=\frac{x_\alpha}{a_\alpha}}$, the 2-metric ${{}^\alpha\mtrt}$ can be rewritten as the atomic metric ${\mtrq_\alpha}$ defined in \eqref{qmtrLCTdef}. Relabeling back ${\xb_\mub\to x_\mub}$, ${\ab_\mub\to a_\mub}$ we get ${{}^{\alpha}\mtrb\to{}^{\alpha{-}1}\mtrc}$.
[Note that relabeling of coordinates does not affect the definition of the rescaled coordinate ${\sx_\mu}$ in the following limiting steps, as ${\sx_\mu = \frac{\xb_\mu}{\ab_\mu}=  \frac{x_\mu}{a_\mu}}$.]
Taking into account relations \eqref{recrel} and the definition \eqref{warplim}, we prove the inductive step
\begin{equation}\label{mtrclimstep3}
\begin{split}
   \mtrc
    &\approx \bigl({}^{N-\alpha}\mtrd + {}^\alpha w^2\; \mtrq_{\alpha}\bigr)
         + {}^\alpha w^2\,\sx_\alpha^2\; {}^{\alpha{-}1}\mtrc\\
    &={}^{N{-}\alpha{+}1}\mtrd + {}^{\alpha{-}1} w^2\; {}^{\alpha{-}1}\mtrc\;,
\end{split}
\end{equation}
and thus the relation \eqref{mtrclimstep1}.
A straightforward check of the last step in the limit shows that after ${N}$ steps we obtain the desired relation ${\mtrc\approx{}^{N}\mtrd=\mtro}$.

For the multiple single-spin-zero limit of the CCKY forms we indicate just few first steps which elucidate the resulting expressions. The relation \eqref{CCKYlimit} for ${\Nt=1}$, together with the same redefinitions as for the metric, allows us to write the first ${k}$ limits as follows
\begin{widetext}
\begin{equation}\label{CCKYmultilimder}
\begin{split}
   \CCKY{k}
   &\approx \sx_N^{2(k{-}1){+}1} \; \LCTq_N\wedge {}^{N{-}1}\!\CCKY{k{-}1}\\
   &\approx \sx_N^{2(k{-}2){+}1}\sx_{N{-}1}^{2(k{-}2){+}1}\; \LCTq_N\wedge\sx_N^2\LCTq_{N{-}1}\wedge
         {}^{N{-}2}\!\CCKY{k{-}2}\\
   &\dots\\
   &\approx \sx_N\dots\sx_{N{-}k+1}\; \LCTq_N\wedge\sx_N^2\LCTq_{N{-}1}\wedge
         \sx_N^2\sx_{N{-}1}^2\LCTq_{N{-}2}\wedge\dots\wedge {}^{N{-}k}\!\CCKY{0}\\
   &={}^k w \; {}^k\LCTd\;.
\end{split}
\end{equation}
The remaining limits do not affect the resulting object any more.
To obtain the limit of KY forms ${\KY{k}}$ we employ the relation \eqref{KYlimit} for the first ${k}$ limits and the relation \eqref{LCTlimit} for the remaining limits:
\begin{equation}\label{KYmultilimder}
\begin{split}
   \KY{k}
   &\approx \sx_N^{2(N{-}k){+}1} \,{}^{N{-}1}\!\KY{k{-}1}\\
   &\dots\\
   &\approx \bigl(\sx_N\dots\sx_{N{+}1{-}k}\bigr)^{2(N{-}k){+}1} \, {}^{N{-}k}\!\KY{0}
      = {}^k\!w^{2(N{-}k){+}1} \;\; {}^{N{-}k}\!\LCT\\
   &\approx {}^k\!w^{2(N{-}k){+}1} \;\; \LCTq_{N{-}k}\wedge\sx_{N{-}k}^{2(N{-}k{-}1)}\;{}^{N{-}k{-}1}\!\LCT\\
   &\dots\\
   &\approx {}^k\!w^{2(N{-}k){+}1} \;\; \LCTq_{N{-}k}\wedge\sx_{N{-}k}^2\,\LCTq_{N{-}k{-}1}\wedge\dots\wedge
      \sx_{N{-}k}^2\dots\sx_{2}^2\,\LCTq_{1} \\
   &={}^k\!w^{2(N{-}k){+}1} \;\; {}^{N{-}k}\LCTo\;.
\end{split}
\end{equation}
The first ${j}$ limits in the multiple single-spin-zero limit of the Killing tensor ${\KT{j}}$ is given by the second relation of \eqref{KTlimit}, the remaining limits act on the lower-dimensional Kerr-NUT-(A)dS metric in a way analogous to \eqref{limmtrc},
\begin{equation}\label{KTmultilimder}
   \KT{j} \approx  {}^{N{-}1}\!\KT{j{-}1}
   \approx \dots
   \approx {}^{N{-}j}\!\KT{0} = {}^{N{-}j}\mtrc^{-1}
   \approx {}^{N{-}j}\mtro^{-1}\;,
\end{equation}
which completes this appendix.
\clearpage
\end{widetext}


%

\end{document}